\documentclass[twocolumn]{aastex63}

\usepackage{xspace}
\usepackage{enumitem}
\usepackage[utf8]{inputenc}
\usepackage{newunicodechar,graphicx}
\usepackage{amsmath}
\usepackage{lineno}

\DeclareRobustCommand{\okina}{%
  \raisebox{\dimexpr\fontcharht\font`A-\height}{%
    \scalebox{0.8}{`}%
  }%
}
\newunicodechar{ʻ}{\okina}

\newcommand{\tess}{\textit{TESS}\xspace}

\newcommand{\kepler}{\textit{Kepler}\xspace}

\newcommand{\tesscut}{\textsf{TESSCut}\xspace}
\newcommand{\lightkurve}{\textsf{lightkurve}\xspace}
\newcommand{\exoplanet}{\textsf{exoplanet}\xspace}
\newcommand{\astropy}{\textsf{astropy}\xspace}
\newcommand{\giants}{\textsf{giants}\xspace}

\newcommand{\hoststar}{TOI-4582\xspace}

\newcommand{\starmass}{$1.34\pm0.06$ $M_\odot$\xspace}
\newcommand{\starradius}{$2.5\pm0.1$ $R_\odot$\xspace}
\newcommand{\teff}{$5190\pm100$ K\xspace}
\newcommand{\feonh}{$0.17\pm0.06$ dex\xspace}

\newcommand{\age}{$4 \pm 1$ Gyr\xspace}
\newcommand{\starrho}{$0.084\pm0.005$ $\rho_\odot$\xspace}
\newcommand{\logg}{$3.77\pm0.04$ dex\xspace}

\newcommand{\planet}{TOI-4582b\xspace}
\newcommand{\planetmass}{$0.53\pm0.05$ $M_\mathrm{J}$\xspace}
\newcommand{\planetradius}{$0.94^{+0.09}_{-0.12}$ $R_\mathrm{J}$\xspace}
\newcommand{\period}{$31.034\pm0.001$ d\xspace}
\newcommand{\transittime}{$1712.177\pm0.004$}

\shorttitle{TESS GTG III: \planet}
\shortauthors{Grunblatt et al.}

\graphicspath{{./}{figures/}}

\begin{document}


\title{\tess Giants Transiting Giants III: An eccentric warm Jupiter supports a period-eccentricity relation for giant planets transiting evolved stars}

\author[0000-0003-4976-9980]{Samuel K. Grunblatt}
\altaffiliation{Kalbfleisch Fellow}
\affiliation{American Museum of Natural History, 200 Central Park West, Manhattan, NY 10024, USA}
\affiliation{Center for Computational Astrophysics, Flatiron Institute, 162 5$^\text{th}$ Avenue, Manhattan, NY 10010, USA}

\author[0000-0003-2657-3889]{Nicholas Saunders}
\altaffiliation{NSF Graduate Research Fellow}
\affiliation{Institute for Astronomy, University of Hawaiʻi at M\=anoa, 2680 Woodlawn Drive, Honolulu, HI 96822, USA}






\author[0000-0003-1125-2564]{Ashley Chontos}
\altaffiliation{Henry Norris Russell Fellow}
\affiliation{Institute for Astronomy, University of Hawaiʻi at M\=anoa, 2680 Woodlawn Drive, Honolulu, HI 96822, USA}
\affiliation{Department of Astrophysical Sciences, Princeton University, 4 Ivy Lane, Princeton, NJ 08544, USA}

\author[0000-0002-0842-863X]{Soichiro Hattori}
\affiliation{Department of Astronomy, Columbia University, 550 West 120$^\text{th}$ Street, New York, NY, USA}

\author[0000-0001-8014-6162]{Dimitri Veras}
\altaffiliation{STFC Ernest Rutherford Fellow}
\affiliation{Centre for Exoplanets and Habitability, University of Warwick, Coventry CV4 7AL, UK}
\affiliation{Centre for Space Domain Awareness, University of Warwick, Coventry CV4 7AL, UK}
\affiliation{Department of Physics, University of Warwick, Coventry CV4 7AL, UK}

\author[0000-0001-8832-4488]{Daniel Huber}
\affiliation{Institute for Astronomy, University of Hawaiʻi at M\=anoa, 2680 Woodlawn Drive, Honolulu, HI 96822, USA}

\author[0000-0003-4540-5661]{Ruth Angus}
\affiliation{American Museum of Natural History, 200 Central Park West, Manhattan, NY 10024, USA}
\affiliation{Center for Computational Astrophysics, Flatiron Institute, 162 5$^\text{th}$ Avenue, Manhattan, NY 10010, USA}
\affiliation{Department of Astronomy, Columbia University, 550 West 120$^\text{th}$ Street, New York, NY, USA}

\author[0000-0002-7670-670X]{Malena Rice}
\altaffiliation{51 Pegasi b Fellow}
\affiliation{Department of Physics and Kavli Institute for Astrophysics and Space Research, Massachusetts Institute of Technology, Cambridge, MA 02139, USA}
\affiliation{Department of Astronomy, Yale University, New Haven, CT 06511, USA}

\author[0000-0001-5228-6598]{Katelyn Breivik}
\affiliation{Center for Computational Astrophysics, Flatiron Institute, 162 5$^\text{th}$ Avenue, Manhattan, NY 10010, USA}

\author[0000-0002-3199-2888]{Sarah Blunt}
\altaffiliation{NSF Graduate Research Fellow}
\affiliation{Cahill Center for Astronomy \& Astrophysics, California Institute of Technology, Pasadena, CA 91125, USA}





\author[0000-0002-8965-3969]{Steven Giacalone}
\altaffiliation{NSF Graduate Research Fellow}
\affiliation{{Department of Astronomy,  University of California Berkeley, Berkeley CA 94720, USA}}

\author[0000-0001-8342-7736]{Jack Lubin}
\affiliation{Department of Physics \& Astronomy, University of California Irvine, Irvine, CA 92697, USA}



\author[0000-0002-0531-1073]{Howard Isaacson}
\affiliation{{Department of Astronomy,  University of California Berkeley, Berkeley CA 94720, USA}}
\affiliation{Centre for Astrophysics, University of Southern Queensland, Toowoomba, QLD, Australia}

\author[0000-0001-8638-0320]{Andrew W.\ Howard}
\affiliation{Cahill Center for Astronomy \& Astrophysics, California Institute of Technology, Pasadena, CA 91125, USA}



\author[0000-0002-5741-3047]{David R. Ciardi}
\affiliation{Caltech/IPAC-NASA Exoplanet Science Institute Pasadena, CA, USA}

\author[0000-0003-1713-3208]{Boris S. Safonov}
\affiliation{Sternberg Astronomical Institute, M.V. Lomonosov Moscow State University, 13, Universitetskij pr., 119234, Moscow, Russia}

\author[0000-0003-0647-6133]{Ivan A. Strakhov}
\affiliation{Sternberg Astronomical Institute, M.V. Lomonosov Moscow State University, 13, Universitetskij pr., 119234, Moscow, Russia}

\author[0000-0001-9911-7388]{David W. Latham}
\affiliation{Center for Astrophysics $\vert$ Harvard \& Smithsonian, 60 Garden St., Cambridge, MA 02138, USA}

\author[0000-0001-6637-5401]{Allyson Bieryla}
\affiliation{Center for Astrophysics $\vert$ Harvard \& Smithsonian, 60 Garden St., Cambridge, MA 02138, USA}

\author[0000-0003-2058-6662]{George R.\ Ricker}
\affiliation{Department of Physics and Kavli Institute for Astrophysics and Space Research, Massachusetts Institute of Technology, Cambridge, MA 02139, USA}

\author[0000-0002-4715-9460]{Jon M.\ Jenkins}
\affiliation{NASA Ames Research Center, Moffett Field, CA, 94035}

\author[0000-0002-1949-4720]{Peter Tenenbaum}
\affiliation{SETI Institute, Mountain View, CA 94043}
\affiliation{NASA Ames Research Center, Moffett Field, CA, 94035}







\author[0000-0002-1836-3120]{Avi~Shporer}
\affiliation{Department of Physics and Kavli Institute for Astrophysics and Space Research, Massachusetts Institute of Technology, Cambridge, MA 02139, USA}

\author[0000-0003-1447-6344]{Edward~H.~Morgan} 
\affiliation{Department of Physics and Kavli Institute for Astrophysics and Space Research, Massachusetts Institute of Technology, Cambridge, MA 02139, USA}

\author[0000-0001-9786-1031]{Veselin Kostov}
\affiliation{NASA Goddard Space Flight Center, 8800 Greenbelt Rd, Greenbelt, MD 20771, USA}
\affiliation{SETI Institute, 189 Bernardo Ave, Suite 200, Mountain View, CA 94043, USA}

\author[0000-0002-4047-4724]{Hugh~P.~Osborn}
\affiliation{Department of Physics and Kavli Institute for Astrophysics and Space Research, Massachusetts Institute of Technology, Cambridge, MA 02139, USA}
\affiliation{NCCR/Planet-S, Universität Bern, Gesellschaftsstrasse 6, 3012 Bern, Switzerland}

\author[0000-0003-2313-467X]{Diana Dragomir}
\affiliation{Department of Physics and Astronomy, University of New Mexico, 210 Yale Blvd NE, Albuquerque, NM 87106, USA}

\author[0000-0002-6892-6948]{Sara Seager}
\affiliation{Department of Physics and Kavli Institute for Astrophysics and Space Research, Massachusetts Institute of Technology, Cambridge, MA 02139, USA}
\affiliation{Department of Earth, Atmospheric, and Planetary Sciences, Massachusetts Institute of Technology, 77 Massachusetts Ave., Cambridge, MA 02139, USA}
\affiliation{Department of Aeronautics and Astronautics, Massachusetts Institute of Technology, 77 Massachusetts Ave., Cambridge, MA 02139, USA}

\author[0000-0001-6763-6562]{Roland K.\ Vanderspek}
\affiliation{Department of Physics and Kavli Institute for Astrophysics and Space Research, Massachusetts Institute of Technology, Cambridge, MA 02139, USA}

\author[0000-0002-4265-047X]{Joshua N.\ Winn}
\affiliation{Department of Astrophysical Sciences, Princeton University, 4 Ivy Lane, Princeton, NJ 08544, USA}






\begin{abstract}

The fate of planets around rapidly evolving stars is not well understood. Previous studies have suggested that relative to the main sequence population, planets transiting evolved stars ($P$ $<$ 100 d) tend to have more eccentric orbits. Here we present the discovery of TOI-4582 b, a \planetradius, \planetmass planet orbiting an intermediate-mass subgiant star every 31.034 days. We find that this planet is also on a significantly eccentric orbit ($e$ = 0.51 $\pm$ 0.05). We then compare the population of planets found transiting evolved (log$g$ $<$ 3.8) stars to the population of planets transiting main sequence stars. We find that the rate at which median orbital eccentricity grows with period is significantly higher for evolved star systems than for otherwise similar main sequence systems, particularly for systems with only one planet detected. In general, we observe that mean planet eccentricity $<e>$ = $a$ + $b$log$_{10}$($P$) for the evolved population with a single transiting planet where $a$ = (-0.18 $\pm$ 0.08) and $b$ = (0.38 $\pm$ 0.06), significantly distinct from the main sequence planetary system population. This trend is seen even after controlling for stellar mass and metallicity. These systems do not appear to represent a steady evolution pathway from eccentric, long-period planetary orbits to circular, short period orbits, as orbital model comparisons suggest inspiral timescales are uncorrelated with orbital separation or eccentricity. Characterization of additional evolved planetary systems will distinguish effects of stellar evolution from those of stellar mass and composition.

\end{abstract}

\section{Introduction} \label{sec:intro}








Though exoplanets have been known around evolved stars at orbital separations $>$1 AU for decades \citep[e.g.][]{hatzes2003}, planets at smaller separations around evolved stars were expected to be engulfed due to angular momentum exchange through tides \citep{hut1981,villaver2009}. In this manuscript, we consider a star to be evolved if its surface gravity $g$ is less than one fourth of the surface gravity of the Sun (i.e., $g$ $\lesssim$ 6,300 cm s$^{-2}$, or log($g$) $<$ 3.8). Until recently, though over 100 planets were known on orbits larger than 1 AU around evolved stars as defined here, no planets were known on orbits smaller than 0.5 AU around evolved stars \citep{schlaufman2013, villaver2014, reffert2015}.

The discovery of Kepler-91b, a Jupiter-sized planet orbiting a 6.3 R$_\odot$, log($g$) $<$ 3.0 star every 6.25 days \citep{lillobox2014,barclay2015}, and subsequent discoveries with the NASA \emph{Kepler} Mission and its extension, \emph{K2} \citep{borucki2010, howell2014}, have proven planets can survive at short periods around evolved stars \citep{huber2013b, almenara2015, vaneylen2016,chontos2019}. These short-period, Jupiter-sized planets have proven useful for studying planet inflation and star-planet interaction, revealing that planets can become inflated at late times \citep{lopez2016, grunblatt2016, grunblatt2017}. Studies of planet occurrence have revealed that these planets are more common than predictions suggested \citep[e.g.,][]{schlaufman2013}, with hot Jupiters being found to be similarly common around main sequence and evolved stars \citep{grunblatt2019}. 

In addition, planets transiting evolved stars have been found to reside on more eccentric orbits, on average \citep{jones2018, grunblatt2018}. Stellar evolution is predicted to enhance the population of closer-in ($\lesssim 0.1 AU$) planets on moderately eccentric orbits, as stellar tides begin to dominate orbital dynamics at larger and larger separations and cause planetary inspiral and orbit circularization simultaneously, producing a transient population of moderately eccentric planets on short periods around evolved stars \citep{villaver2014}. However, this cannot explain why these planets had larger ($> 0.1 AU$), highly-eccentric orbits during the main sequence.

These large main sequence eccentricities can be potentially explained by planet-planet scattering events in the initial stages of planetary system formation and evolution. Stars with higher masses or higher metallicities might have formed in conditions that led to multiple closely-spaced giant planets, which in turn could have led to planet-planet scattering events, known to excite orbit eccentricities \citep{frelikh2019}. In addition, evolved transiting planetary system host stars are more massive and metal-rich than the Sun, on average \citep{grunblatt2019}. 

Unfortunately, the majority of evolved transiting planetary systems currently known cannot constrain the rate of planet-planet scattering events in these systems due to the possibility of strong tidal dissipation during hot Jupiter formation, which effectively erases the evidence for the formation pathways of hot Jupiter systems \citep{dawson2018}. In contrast, planets at larger orbital separations ($>$ 0.1 AU) do not experience such strong tidal dissipation, and thus may effectively ``fossilize'' evidence for  planet-planet scattering by maintaining high eccentricities at late evolutionary stages. If planets on longer periods around evolved stars tend to be more eccentric than planets around main sequence stars, this can be interpreted as evidence for higher rates of planet-planet scattering in these systems. Determining whether planet-planet scattering is more prevalent in evolved planetary systems than in equivalent main sequence systems indicates if this scattering is a result of post main sequence system evolution or initial planetary system formation, essential to understanding the long-term stability of planetary systems.

At least one planet has been found transiting an evolved host star with a period longer than 50 days, Kepler-432 b \citep{quinn2015,ciceri2015}. This planet has a particularly high orbital eccentricity ($e$ = 0.54 $\pm$ 0.03). This high eccentricity, as well as evidence for the existence of an additional companion in the system, suggests that Kepler-432 is potentially a dynamically active planetary system. It is unclear what role each of the components of the system has played in producing the currently observed orbital configuration, but planet-planet interactions likely played an important role in sculpting the system architecture. 

Radial velocity studies have also revealed a long-period planet population around evolved stars at separations $>$0.5 AU. These planets have a wide range in eccentricities, but interestingly appear to have generally lower eccentricities than planets of similar mass and orbital separation around main sequence stars \citep{jones2014,grunblatt2018}. This is seemingly at odds with the moderately high eccentricities of planets observed at smaller periods around evolved stars, but may imply that planet-planet scattering is common within 0.5 AU of evolved host stars and rarer at larger orbital separations. As gas giant planets are believed to form beyond 1 AU, this suggests that most transiting planets observed around evolved stars have undergone some sort of planet-planet scattering, while planets detected at larger separations have not experienced as much dynamical interaction. The characterization of more planets around evolved systems with intermediate orbital periods (0.1-1 AU) will reveal if this difference in eccentricity distribution is more dependent on orbital separation or if it is a selection effect, and if a turnover in median eccentricity can be found at longer periods. A search for transiting planets across the entire sky will be essential for detection of these intermediate orbital period planets around evolved stars. 

The Transiting Exoplanet Survey Satellite (\tess; \citealt{ricker2014}) is enabling the discovery of a predicted $\sim$14,000 planets \citep{sullivan2015,barclay2018, kunimoto2022} across the entire sky. During its 2-year Primary Mission (July 2018 - July 2020), the space telescope observed stars in Full Frame Images (FFIs) with a 30-minute observing cadence, and completed one year of observations in each of the northern and southern ecliptic hemispheres. Each year was split into 13 observing sectors that stretched from the ecliptic pole toward the ecliptic plane, moving every $\sim$27 days. Targets near the ecliptic pole were observed in multiple sectors, in some cases providing a full year of photometry, while targets closer to the ecliptic plane were observed in fewer sectors. According to the NASA Exoplanet Science Institute (NExSci) archive\footnote{\href{https://nexsci.caltech.edu}{nexsci.caltech.edu}}, \tess has already led to the discovery of 100+ confirmed planets and 5,000+ project candidates \citep{guerrero2021}. Of the planets confirmed to date, only a handful orbit post main sequence stars, most of which have only just begun evolution off of the main sequence \citep[e.g.,][]{huber2019,nielsen2019,wang2019,eisner2020,rodriguez2021,khandelwal2022,montalto2022,saunders2022,wittenmyer2022, grunblatt2022}.

Here we present analysis of the \tess\ Prime Mission and available spectroscopic and imaging data for \hoststar\, and place this planetary system into context of the known population of giant planets transiting evolved stars. We find that \planet seems to support the existence of a period-eccentricity correlation for planets transiting evolved stars, and that the population of transiting planets around evolved stars is significantly distinct from the population of similar planets transiting main sequence stars in the period-eccentricity plane. We investigate whether this difference could have arisen from other fundamental differences between these populations, such as their stellar masses and metallicities, or if it is more clearly correlated with evolution proxies such as stellar radius or age. We also show how the period-eccentricity relation is relatively insensitive to our definition of `evolved,' and investigate whether this correlation arises from a clear evolutionary pathway from longer period, eccentric orbits to small, circular orbits at late evolutionary stages. Finally, we consider future prospects for investigating the evolution of planetary architectures.

%




\section{Observations} \label{sec:methods}

\subsection{\tess Photometry} \label{sec:photo}

\begin{figure*}[ht!]
    \centering
    \includegraphics[width=0.95\textwidth]{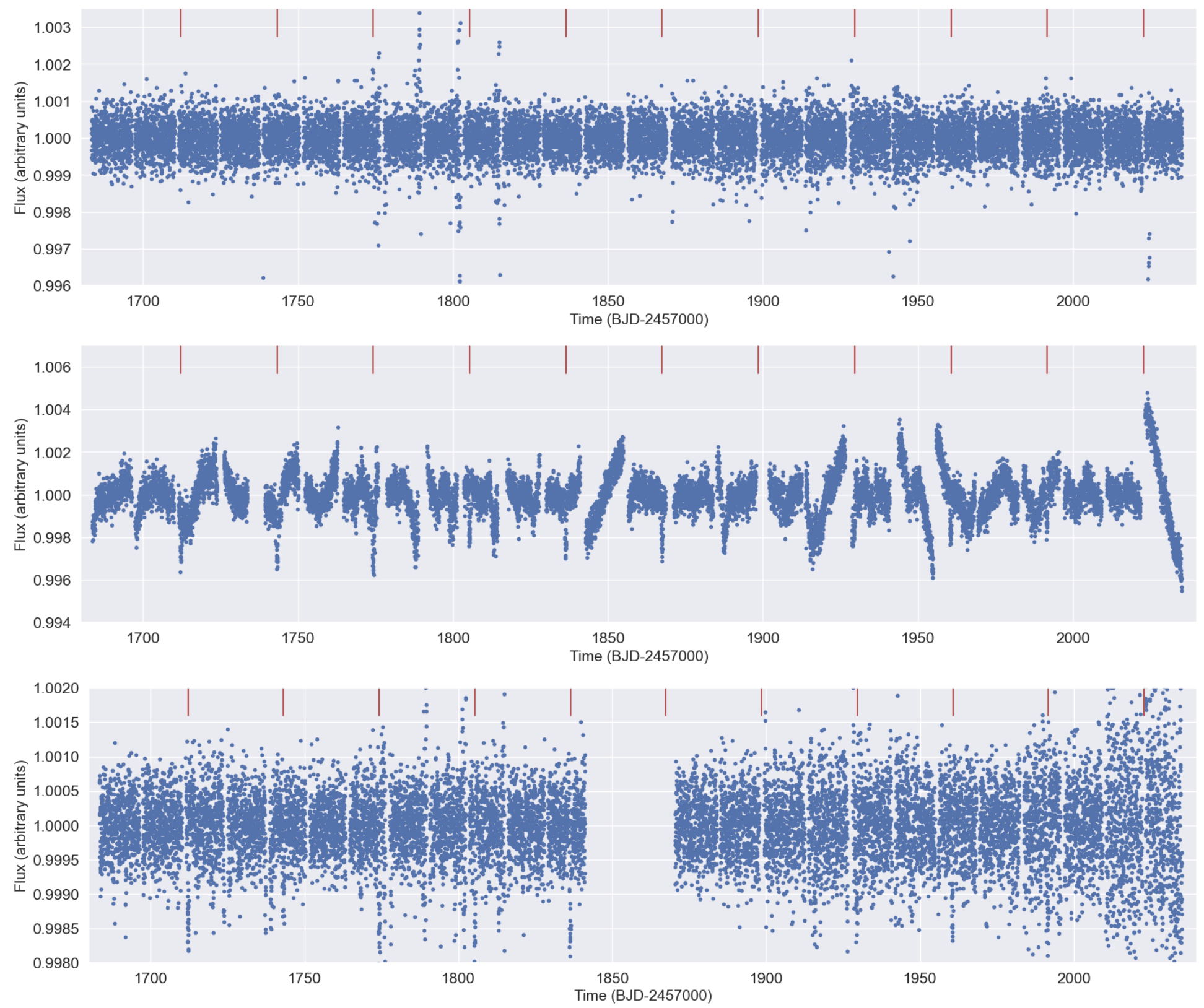}
    \caption{Unfolded, full \tess\ prime mission light curves of \hoststar with 'KSPSAP FLUX', 'SAP FLUX', and \giants\ flux values (from top to bottom). Transits are indicated by red fiducial marks. While the 'KSPSAP FLUX' smoothing removes both transits and long-term trends visible in the 'SAP FLUX' lightcurve, the \giants\ approach to detrending of the light curve removes long-term variability while preserving planet transits, despite their long duration. However, one sector of data around BJD 2458850 is not included in the \texttt{giants} light curve due to poor background subtraction. }
    \label{fig:unfolded_lc}
\end{figure*}

\planet was discovered as part of our survey to identify and confirm planets around evolved stars using \tess FFIs (Guest Investigator programs GO22102,GO3151,GO4179). \hoststar\ was observed by the \tess\ Prime Mission in Sectors 14-26, from July 18, 2019, to July 4, 2020. Using the \tess Input Catalog \citep[TICv8;][]{stassun2019}, we made cuts based on \emph{Gaia} color (B$_p$ - R$_p$ $>$ 0.9 mag), absolute magnitude (M$_G$ $<$ 4.0 mag), and apparent magnitude (m$_T$ $<$ 13 mag) in order to limit our sample to relatively bright, evolved stars. We developed the \giants\footnote{\href{https://github.com/nksaunders/giants}{https://github.com/nksaunders/giants}} Python package for accessing, de-trending, and searching \tess observations for periodic transit signals \citep{saunders2022}. The details of how this pipeline processes \tess\ full frame image data are described in detail in \citet{saunders2022}, but in short, the \texttt{giants} pipeline uses a principal component analysis approach to detrending. In summary, a light curve is produced by creating an 11x11 pixel cutout for all sectors the targets is observed by \tess\, then drawing an aperture on the cutout around the target based on a threshold flux. 10 principal components common to the cutout outside the chosen aperture are then identified and subtracted from the flux within the aperture, leaving the signal that is present within the aperture without significant background contribution. We also present the \giants\ light curves along with the 'SAP FLUX' and 'KSPSAP FLUX' light curves produced by the MIT Quick Look Pipeline \citep{huang2020} in Figure \ref{fig:unfolded_lc}, which use a background subtraction and contamination factor-based approach to produce light curves, where the 'KSPSAP FLUX' light curve also includes a spline-based detrending step. \planet\ is on a long-period, highly eccentric orbit, which was not detected by other early \tess\ transit searches due to its long transit duration.

However, a later transiting planet search by the TESS Science Processing Operations Center \citep[SPOC;][]{jenkins2016} of sectors 40 and 41 on 14 September 2021 detected the signatures of two transits of \planet \citep{jenkins2002,jenkins2010,jenkins2020},  which was fitted with an initial limb-darkened transit model \citep{li2019} and passed all the diagnostic tests \citep{twicken2018}. The TESS Science Office issued an alert for TOI-4582b on 11 November 2021 reviewing the data validation reports \citep{guerrero2021}. A subsequent search by the SPOC of sectors 40, 41 and 46-50 redetected the transit signatures and yielded difference image centroiding results that located the source of the transits to within 1.8 $\pm$ 2.9 arc sec from the host star.



We used our \texttt{giants} pipeline to produce \tess\ light curves for all stars which passed our aforementioned cuts. We produced approximately 540,000 light curves from the first 2 years of data from the \tess\ Mission. We then performed an automated BLS search on these targets, and produced summary plots using the BLS output as well as TIC information and the pixel cut out. These summary plots were then visually inspected, during which \hoststar was flagged for potential rapid ground-based followup. The results of this visual inspection will be verified by computational techniques and a subsequent catalog of planet candidates transiting evolved stars will be released in the near future (Saunders et al., {\it in prep.}). Eleven sectors of data were available for \planet at the conclusion of the \tess\ prime mission, during which nine transits were observed. Additional data is currently being taken for this target as part of the \tess\ extended mission. 

\begin{figure}[h!]
    \centering
    \includegraphics[width=0.495\textwidth]{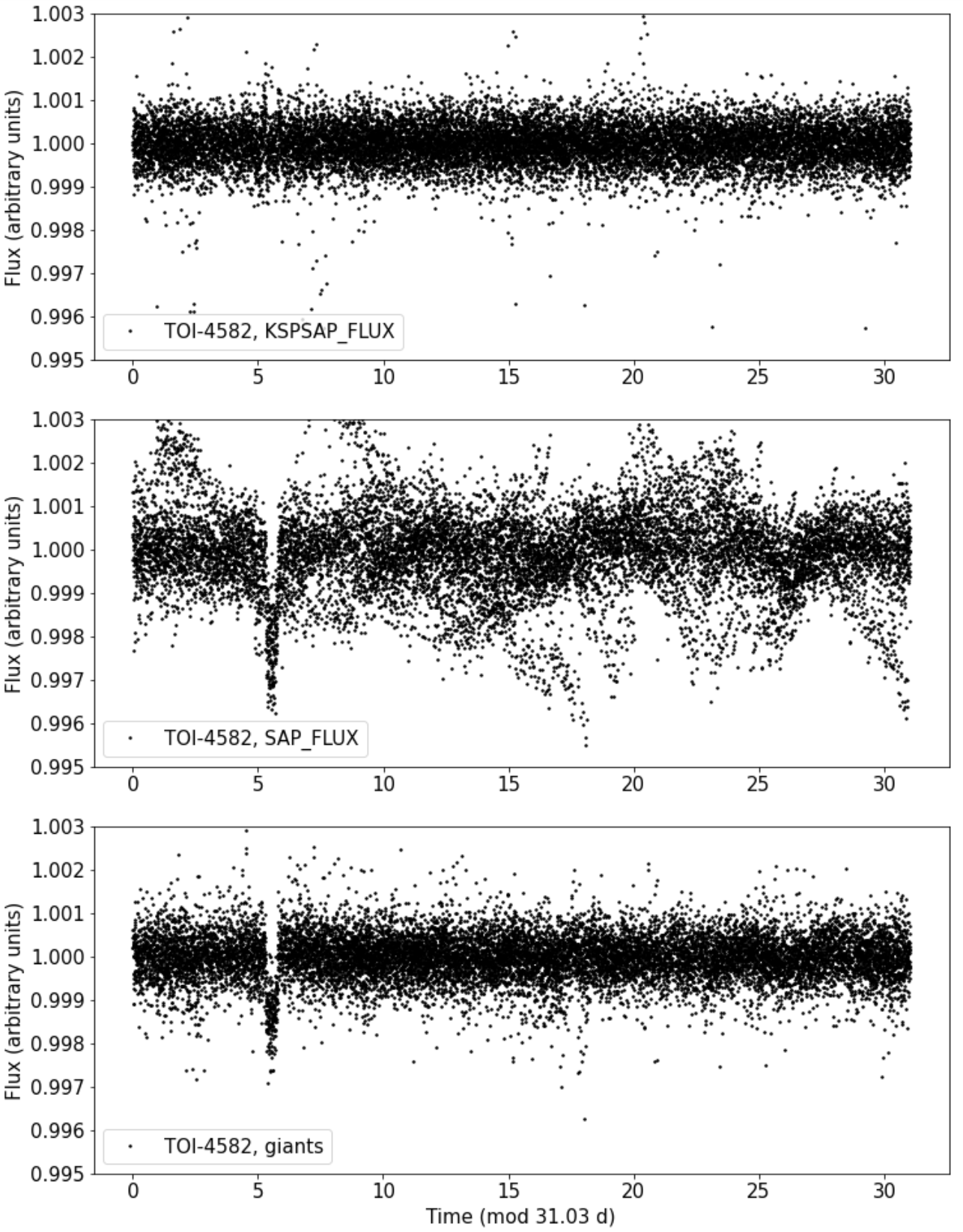}
    \caption{Phase-folded light curves of \hoststar using the QLP `KSPSAP FLUX', `SAP FLUX', and \texttt{giants} fluxes (from top to bottom, respectively). Clear differences in transit depth and shape and light curve scatter can be seen between the different light curves.}
    \label{fig:2567_comp}
\end{figure}

We illustrate the phase-folded light curves of \hoststar in Figure \ref{fig:2567_comp}. The planet transit can be seen at a phase of 6 days in the `SAP FLUX' and \giants\ light curves. The transit duration has been measured to be 11.952 $\pm$ 1.368 hrs. As the `KSPSAP FLUX' pipeline is smoothed using a spline approach with a window length of 0.3 days, it is likely that the transit of \planet was smoothed to the point where it is no longer visible. This hypothesis is supported by the lack of transit seen in the phase-folded `KSPSAP FLUX' light curve while transits are clearly seen in the `SAP FLUX' and \giants\ light curves. The principal component analysis approach to detrending used by the \giants\ pipeline removes trends which are seen in neighboring \tess\ pixels, thus reducing the effect of scattered light and \tess\ orbital properties while preserving the transit signal, which is detected only in the target pixels. This approach can fail in sectors where background light is sufficiently strong both in and outside the chosen aperture for the target, here resulting in the data gap around BJD $\sim$ 2458850 seen in the \texttt{giants} light curve in Figure \ref{fig:unfolded_lc}.

In addition, a significant difference in transit depth is measured between the `SAP FLUX' and \giants\ light curves, where the `SAP FLUX' transit is $\approx$15\% deeper than the \giants\ transit, implying a planet radius which is $\approx$8\% larger. This difference and its implications are discussed further in Section \ref{sec:planetprops}.

\subsection{Radial Velocity Measurements}

RV observations were taken with the HIRES spectrograph on the Keck-I telescope on Maunakea, Hawaii \citep{vogt1994}. HIRES has a resolving power of $R\approx60,000$ and wavelength coverage between $\sim350$nm and $\sim620$nm. 12 RV measurements were taken of \hoststar between 2021 May 27 and October 17. We list our RV measurements and uncertainties in Table \ref{table:rvs}.

In addition, 2 RV observations were taken between November 2021 and March 2022 using the TRES spectrograph on the 1.5m FLWO telescope \citep{furesz2008}. These measurements revealed a radial velocity offset of 99 m s$^{-1}$ that is in phase with the photometric ephemeris and is consistent with a planetary-mass companion. Given the higher precision and phase coverage of Keck/HIRES, we only list the Keck/HIRES RV measurements here.

\begin{table}[]
    \centering
    \begin{tabular}{c c}
        \hline
        Time (JD) & Relative RV (m/s)  \\
        \hline
        2459361.983 & 30.9 $\pm$ 0.8 \\
        2459376.920 & -24.3 $\pm$ 1.0 \\
        2459385.849 & -11.7 $\pm$ 1.6 \\
        2459399.906 & -3.5 $\pm$ 1.4 \\
        2459406.904 & -11.9 $\pm$ 1.0 \\
        2459435.825 & -21.6 $\pm$ 1.4 \\
        2459441.914 & -17.1 $\pm$ 1.4 \\
        2459451.859 & 26.0 $\pm$ 1.5 \\
        2459456.838 & 21.6 $\pm$ 1.5 \\
        2459475.790 & -7.6 $\pm$ 1.7 \\
        2459484.773 & 44.7 $\pm$ 1.6 \\
        2459504.787 & -25.0 $\pm$ 1.9 \\
        \hline
    \end{tabular}
    \caption{Radial velocities and uncertainties measured for \hoststar from Keck/HIRES. The RVs have been sorted in time.}
    \label{table:rvs}
\end{table}

\section{Host Star Characterization} \label{sec:hoststar}


\begin{figure}[ht!]
    \centering
    \includegraphics[width=.5\textwidth]{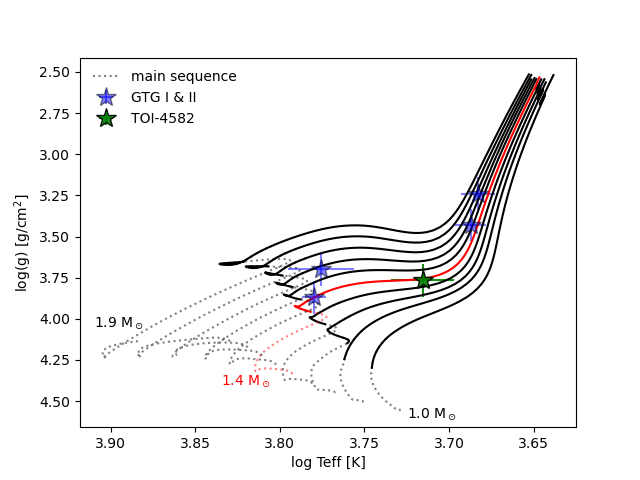}
    \caption{Position of \hoststar along with the other planet hosts from the Giants Transiting Giants program shown on an H-R diagram. All host stars have evolved off of the main sequence onto the subgiant and red giant branch. We also illustrate MIST evolutionary tracks of 1-2 M$_\odot$, +0.25 [Fe/H] dex stars in 0.1 M$_\odot$ increments for reference. We have highlighted a MIST evolutionary track for a 1.4 M$_\odot$, [Fe/H] = 0.25 dex star in red, illustrating the evolutionary sequence directly probed here.}
    \label{fig:hr}
\end{figure}

\begin{table*}
\centering
    \begin{tabular}{l c c c}
        \hline
        \rule{0pt}{3ex}\textit{Target IDs} & & & \\
        \rule{0pt}{3ex}TOI & 4582 &  & \\
        TIC & 219854519 &  &  \\
        TYC & 4421-01472-1 & &  \\
        2MASS & J17072649+6851562 & &  \\
        Gaia DR2 & 1636984973365053056 &  & \\
        \hline
        \rule{0pt}{3ex}\textit{Coordinates} & \\
        \rule{0pt}{3ex}RA(J2015.5) & 17:07:26.43  &  & \\
        Dec(J2015.5) & $+$68:51:56.25 &  & \\
        \hline
        \rule{0pt}{3ex}\textit{Characteristics} & \\
        \rule{0pt}{3ex}\tess\ mag & 10.54 &  & \\
        Radius $R_\star$  & \starradius &  \\
        Mass $M_\star$  & \starmass & \\
        $T_{\rm eff}$  $ $ & \teff &  \\
        $\log(g)$  & \logg &  \\
        $ $[Fe/H]  $ $ & \feonh &  \\
        $v$sin$i$  $ $ & 2.7 $\pm$ 1.0 km s$^{-1}$&  \\
        Age  $ $ & \age &  \\
        Density $\rho_\star$  & \starrho  \\
        \hline
   \end{tabular}
	 \caption{Effective temperature, surface gravity and metallicity were determined using \texttt{SpecMatch-Syn} fits to Keck/HIRES spectra. Stellar mass, radius, age, and density are derived from an \textsf{isoclassify} fit to HIRES spectroscopic observations, \emph{Gaia} parallaxes and 2MASS K-band photometry. We inflate our error bars reported from isochrones to reflect more realistic uncertainties following the reasoning of \citet{tayar2022}.}
	 \label{table:stellar}
\end{table*}

\subsection{High-Resolution Spectroscopy}

We used \texttt{SpecMatch-Syn} to measure the metallicity, surface gravity and effective temperature of the host star \hoststar from our HIRES spectra \citep{petigura2015}. We then used \textsf{isoclassify} \citep{huber2017} to combine \emph{Gaia} parallax measurements with spectroscopic information to determine stellar properties, listed in Table \ref{table:stellar}. We find that these results are in good agreement with an independent analysis of stellar parameters using the SPC (stellar parameter classification) technique \citep{buchhave2012}, which follows a similar general procedure as \texttt{isoclassify} but use different stellar models to describe spectral observations of this target with the TRES spectrograph \citep{furesz2008}. SPC and \texttt{SpecMatch-Syn} both cross-correlate an observed spectrum against a grid of synthetic spectra, and use the correlation peak heights to fit a three-dimensional surface in order to find the best combination of atmospheric parameters. 

In this case, the SPC analysis did not use priors to constrain stellar parameters based on a grid of isochrones, as the overlap of isochrone grids of different metallicities on the subgiant branch used by SPC can result in biased determinations of stellar surface gravities. As the subgiant stage of evolution is relatively short, fitting to isochrone grids can provide very precise parameter constraints for subgiant stars. However, fine tuning of stellar model parameters that are poorly constrained such as mixing length can result in substantial changes in the inferred intrinsic stellar parameters of stars on the subgiant branch. Thus, we report errors which are modestly larger than what is returned from our isochrone-based analysis to more accurately reflect realistic uncertainties on the intrinsic stellar parameters, inflating errors to match the known fractional systematic uncertainty floor for stars at this evolutionary state \citep{tayar2022}.




Figure \ref{fig:hr} shows an H-R diagram with evolutionary tracks downloaded from the MESA Isochrones \& Stellar Tracks (MIST; \citealt{dotter2016}; \citealt{choi2016}; \citealt{paxton2011}). We have shown the positions of all host stars of confirmed planets found by our \emph{TESS} Giants Transiting Giants Guest Investigator program. As all five host stars have roughly the same mass and metallicity (M$_*$ $\approx$ 1.5 M$_\odot$, [Fe/H] $\approx$ 0.25 dex), we suggest that these systems may represent an evolutionary sequence for post-main sequence, intermediate-mass stars. We find that \hoststar lies at a late subgiant stage of evolution between the evolutionary stage of subgiants TOI-2184 and TOI-4329, and giant stars TOI-2337 and TOI-2669. 



\subsection{Rotation of \hoststar}

Outside of the transit, the `SAP FLUX' light curve of \hoststar shows smooth, longer-period variability which may be associated with the rotation of the host star. Measurement of the stellar rotation period can constrain strength of magnetic activity and tidal interaction with the orbiting planet. Furthermore, although stellar rotation periods cannot yet be used to constrain the age of subgiant stars, as stellar rotation rates have been directly measured for only a handful of subgiants, a measurement of the stellar rotation rate in this system may provide information about the correlation between age and rotation rates of subgiants in the future, and thus is worth reporting if it can be measured using reliable methods. In order to produce a light curve particularly sensitive to rotational variation, we use the \texttt{unpopular} package to perform causal pixel modeling and isolate the long-period variability of this target \citep{hattori2021}. We find a tentative rotation period detected at P $\sim$ 73 d, with different choices of detrending parameters resulting in rotation periods detected between 71 and 77 days, in agreement with rotation rates and signal amplitudes reported by \citet{santos2021} for similar subgiant targets. Confirmation of this signal with other high cadence photometric surveys, such as the Zwicky Transient Facility \citep{bellm2019}, will test the validity and accuracy of this stellar rotation period measured here.

\subsection{Constraints on Binarity of \hoststar} 

\begin{figure}[ht]
    \centering
    \includegraphics[width=.5\textwidth]{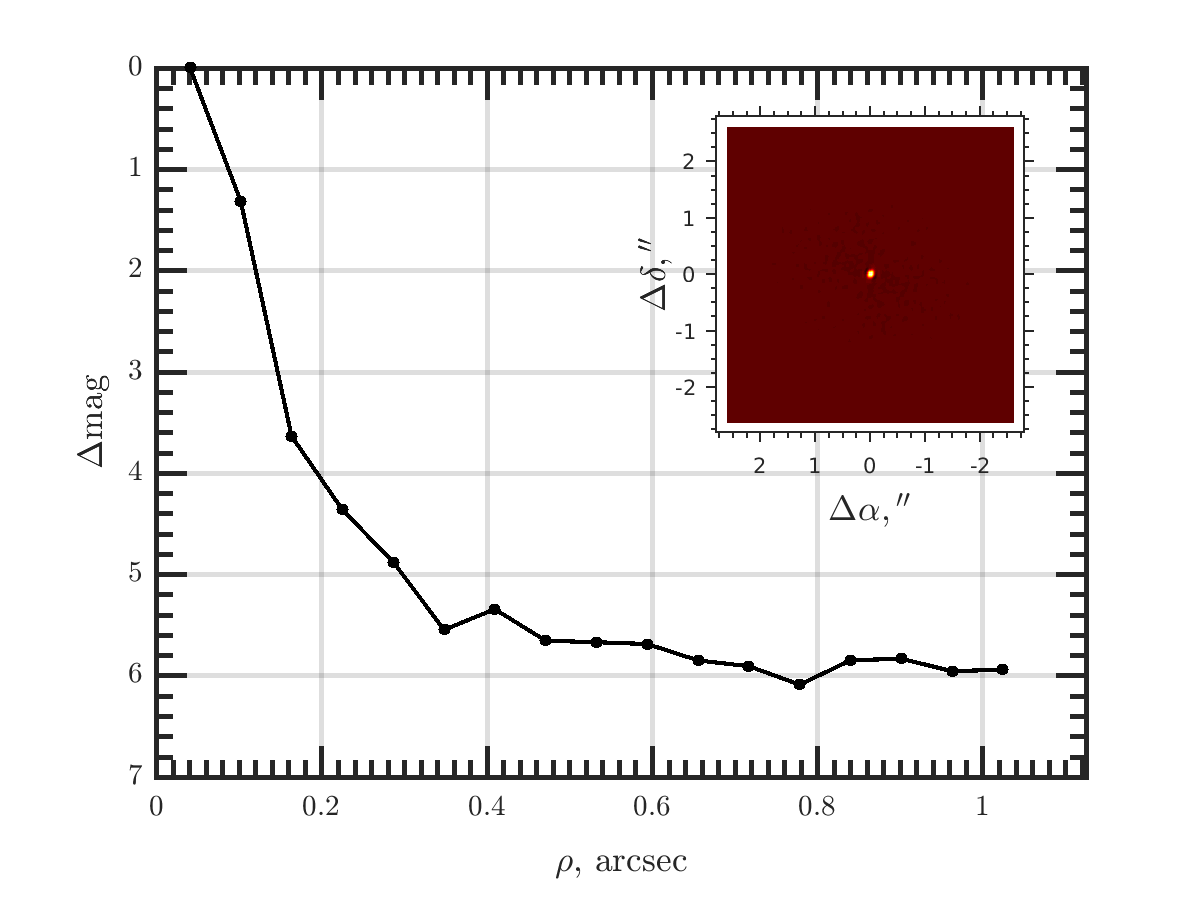}
    \caption{Contrast curve for \hoststar based on speckle polarimeter imaging taken with the Sternberg Astronomical Institute 2.5m telescope, with the corresponding speckle autocorrelation function shown in the inset. \hoststar appears to be a single star based on imaging data from multiple sources and astrometric information from \emph{Gaia}. }
    \label{fig:imaging}
\end{figure}

\begin{table*}
\begin{center}
    \begin{tabular}{l c r}
        \hline 
        Parameter & Prior & Value \\
        \hline
        \rule{0pt}{3ex}\textit{Transit Fit Parameters} & & \\
        \rule{0pt}{3ex}Orbital period $P_{\text{orb}}$ [days] & $\log\mathcal{N}[31.06, 0.0015]$ &  \period \\
        Planet-to-star radius ratio $R_p/R_
        *$ & $\mathcal{U}[0, 1]$& 0.0351 $\pm$ 0.002 \\
        Transit epoch $t_0$ [BJD - 2457000, TDB] & $\mathcal{N}[1712.0, 1.0]$ & \transittime \\
        Impact parameter $b$ & $P_\beta(e\in[0,1])^\text{(a)}$  & $0.52\pm0.08$\\
        Eccentricity $e$ & single-planet dist. from \citet{vaneylen2019} & $0.51 \pm 0.05$\\
        Argument of periastron $\Omega$ & $\mathcal{U}[-\pi, \pi]$& -0.166 $\pm$ 0.102 \\
        Limb-darkening coefficient $q_1$ & [0,2]$^\text{(b)}$ & $0.383 \pm 0.224$ \\
        Limb-darkening coefficient $q_2$ & [-1,1]$^\text{(b)}$ & $0.176 \pm 0.327$ \\
        \hline
        \rule{0pt}{3ex}\textit{Radial Velocity Fit Parameter} & & \\
        \rule{0pt}{3ex}Semi-amplitude $K$ [m/s] & $\mathcal{U}[0, 500]$ & $31 \pm 3$\\
        \hline
        \rule{0pt}{3ex}\textit{Derived Physical Parameters} & & \\
        \rule{0pt}{3ex}Planet radius $R_p$ $(R_J)$ & $\mathcal{U}[0, 3]$ & 0.94$^{+0.09}_{-0.12}$ \\
        Planet mass $M_p$ $(M_J)$ & $\mathcal{U}[0, 300]$ & 0.53 $\pm$ 0.05 \\
        \hline
   \end{tabular}
	 \caption{Fit and derived parameters for \planet. \textit{Note:} $^\text{(a)}$This parameterization is described by the Beta distribution in \citet{kipping2013b}. $^\text{(b)}$  Distributions follow correlated two-parameter quadratic limb-darkening law from \citet{kipping2013}.}
	 \label{table:planet}
\end{center}
\end{table*}

\hoststar was observed with the Speckle Polarimeter \citep[SPP;][]{safonov2017} on the 2.5-m telescope at the Caucasian Observatory of the Sternberg Astronomical Institute on March 14th, 2022 using the Ic-band filter near 0.9 $\mu$m. SPP uses an Electron Multiplying CCD Andor iXon 897 as a detector. The atmospheric dispersion compensator was employed. The detector has a pixel scale of 20.6 mas/pixel, and the angular resolution of the observations is 89 mas, with a field of view of 5'' $\times$ 5'' centered on \hoststar. The power spectrum was estimated from 4000 frames with 30 ms exposure. The contrast curve for the SAI observations can be seen in Figure \ref{fig:imaging}, which shows the detection limits in contrast ($\Delta$m) versus angular separation from the point spread function center in arcseconds for the filter wavelength. The inset image is the speckle auto-correlation function for the observation. We did not detect any stellar companions brighter than $\Delta$m = 4.0 and 5.6 at 0.2'' and 0.5'', respectively. \hoststar was also observed by the NESSI instrument on the 3.5m WIYN telescope at the Kitt Peak Observatory in Arizona \citep{scott2018}. Observations were taken on April 21st, 2022 using the 832 nm filter.  No additional stars can be identified within 4'' of our target in either imaging dataset. The difference image centroiding results produced by SPOC also agree with the results of the high resolution imaging.

Furthermore, the Gaia astrometric noise metric RUWE for \hoststar is low (0.77), indicative that the star is not in a wide binary system which could be resolved by \emph{Gaia} astrometry (typical RUWE values for binaries are $>$1.4). Additionally, no evidence of a spectroscopic binary can be seen in the spectra of this star, placing limits on a companion of similar brightness for this star. Specifically, we use the \citet{kolbl2015} routine to determine limits on spectroscopic binarity, and find no evidence for a spectroscopic companion brighter than 1\% of the brightness of the primary with a radial velocity shift of 10 km/s or larger from the primary. The radial velocity measurements of these systems do not show any significant linear or quadratic trends with time, suggesting this star is single and not part of a binary system.

\section{Planet Characterization} \label{sec:planetprops}

\begin{figure*}[ht]
    \centering
    \includegraphics[width=.46\textwidth]{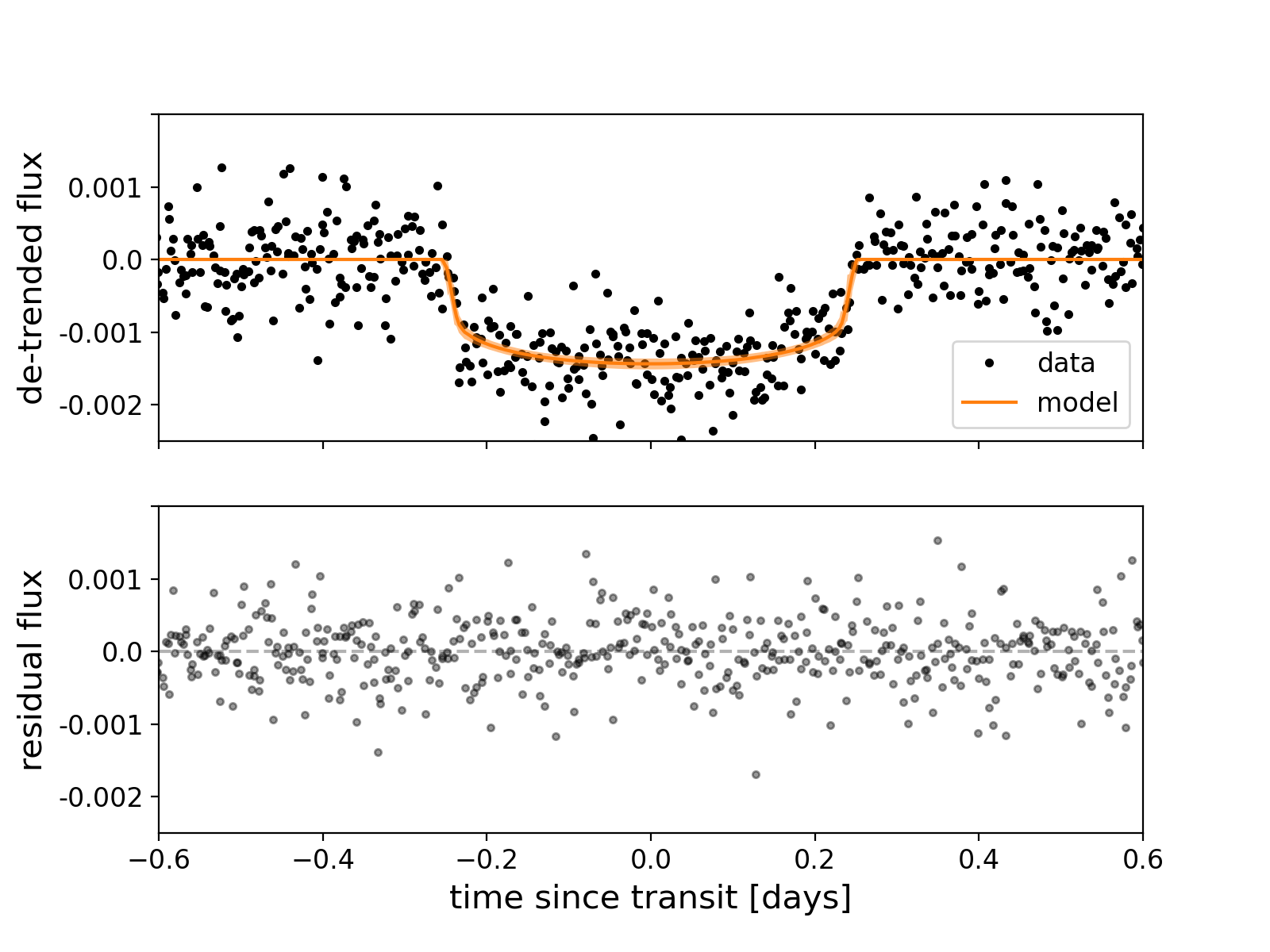}
    \includegraphics[width=.49\textwidth]{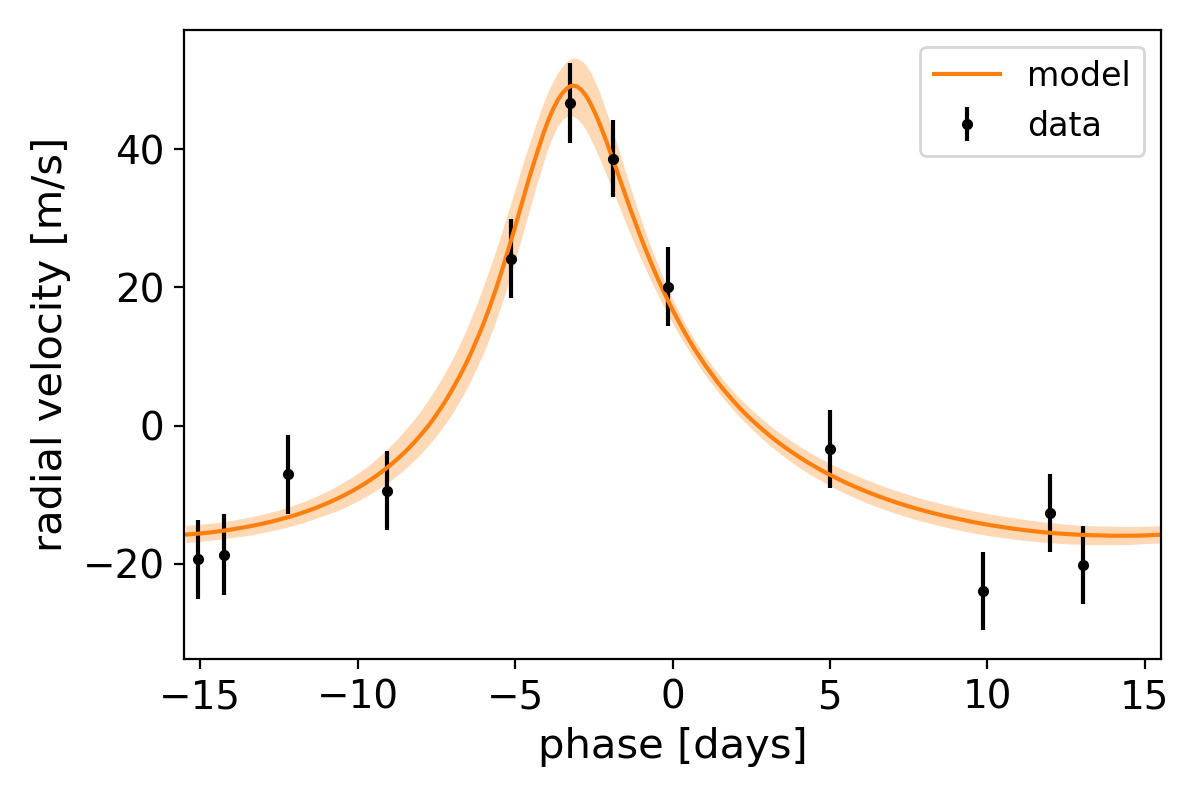}
    \caption{{\it Left:} The light curve of \hoststar folded at a period of \period days. The de-trended photometry is shown in black with the transit model from \texttt{exoplanet} has been overplotted in orange. The residuals to the model are shown in the lower panel. {\it Right:} All Keck/HIRES radial velocity observations of \planet along with the best-fit \texttt{exoplanet} model used in this analysis, where the time axis has been folded at the orbital period of the planet. Radial velocity errors include contribution from both measurement and a white noise term within the model.}
    \label{fig:lc_rv_2300}
\end{figure*}

\subsection{Model Fit}

We used the \exoplanet Python package to simultaneously fit a model to the photometry and radial velocity observations \citep{exoplanet:exoplanet}. The data input to our model included all Keck/HIRES radial velocity observations and all sectors of \tess FFI photometry available from the first three years of the \tess\ Mission. Our model used the stellar parameters given in Table \ref{table:stellar}. 

Our initial choices of planet period and depth were taken from the BLS search determined values produced during the transit search described in \S 2.1. For limb darkening, we use the quadratic model prescribed by \cite{exoplanet:kipping13} to provide a two-parameter model with uninformative sampling. We parameterized eccentricity using the single planet eccentricity distribution of \citet{vaneylen2019}. We present our best fit models to the \giants\ light curve and radial velocity data for \hoststar in Figure \ref{fig:lc_rv_2300} and Table \ref{table:planet}.  

We also performed a joint radial velocity and transit fit using the `SAP FLUX' light curve (in addition to the \giants\ light curve). We find a planet radius of 0.88 $\pm$ 0.06 R$_\mathrm{J}$ using the \giants\ light curve, and 1.01 $\pm$ 0.03 R$_\mathrm{J}$ using the `SAP FLUX' light curve. The range in planet radii determined by these fits is 0.13 R$_\mathrm{J}$, significantly larger than the uncertainty in planet transit depth estimated from any fit to the available data. We note that this discrepancy in transit depth between the \giants\ pipeline and other pipelines has been seen in other transiting planet examples \citep[such as TOI-4329,][]{grunblatt2022} and may be related to background light contamination or baseline flux determination within the multipixel aperture, which the \giants\ pipeline does not account for in the same fashion as the 'SAP FLUX' or 'KSPSAP FLUX' pipelines. We note that an independent analysis of the short-cadence SPOC light curve produced by the \tess\ Extended Mission of this target successfully detects this transit and measures an $R_p$/$R_*$ = 0.0352 for this system, almost identical to the value reported by our \giants\ pipeline analysis. Despite this, we choose to keep our error estimates conservative, and thus we adopt the full range of radius uncertainties from both light curves as our final reported uncertainties and the mean transit radius between the \giants\ and `SAP FLUX' light curve fits, and report a planet radius of \planetradius for \planet. Future observations with facilities with better resolution capable of using smaller photometric apertures, such as those used by the Las Cumbres Observatory Global Telescope array \citep{brown2013}, will be able to confirm the true transit depth of this system, more precisely constraining the radius of \planet. Despite the discrepancy in radius measurements using different \tess\ FFI light curves, our main conclusions about the system and its context in the larger known planet population are unaffected and would remain valid if the planet radius measured exclusively from either the 'SAP FLUX' or \texttt{giants} light curve were used for this study.

\begin{figure*}[ht!]
    \centering
    \includegraphics[width=\textwidth]{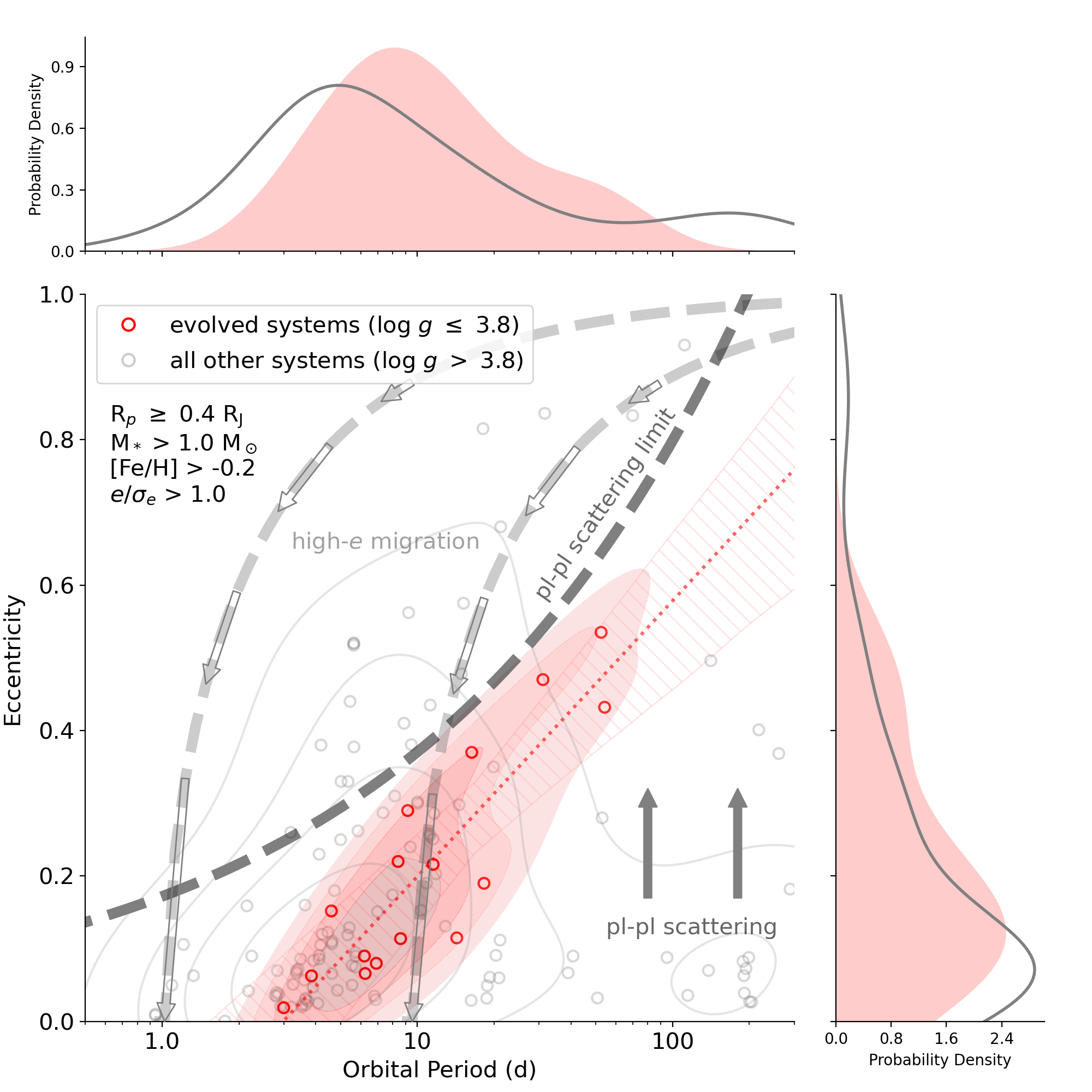}
    \caption{Eccentricity as a function of orbital period for planets in systems where R$_p$ $\geq$ 0.4 R$_\mathrm{J}$, orbiting stars where $M_* > 1.05 M_\odot$ with metallicities $>$ -0.2 dex, with significant eccentricities. Systems transiting evolved (log($g$) $\geq$ 3.8) stars have been shown by the red contours and corresponding red points, while planets transiting main sequence (log($g$) $>$ 3.8) stars are shown by the gray contours and points. Tracks of constant orbital angular momentum followed during high-eccentricity migration are shown by the light gray dotted lines and arrows, while the effects of planet-planet scattering and its approximate limit determined following the formulation noted in \citet{dawson2018} are shown by the dark gray arrows and dotted line. We find that planets transiting evolved stars appear to form a more linear distribution in period-eccentricity space than the rest of the planet population, possibly sculpted by these migration and scattering processes. Furthermore, we find that the orbital eccentricity of evolved systems can be approximated well by a linear regression to the logarithm of the orbital period, shown by the red dotted line and hatched region corresponding to a 95\% confidence interval. A similar linear correlation is significantly weaker for the overall planet population.}
    \label{fig:perecc_v1}
\end{figure*}

\begin{figure}[ht!]
    \centering
    \includegraphics[width=.49\textwidth]{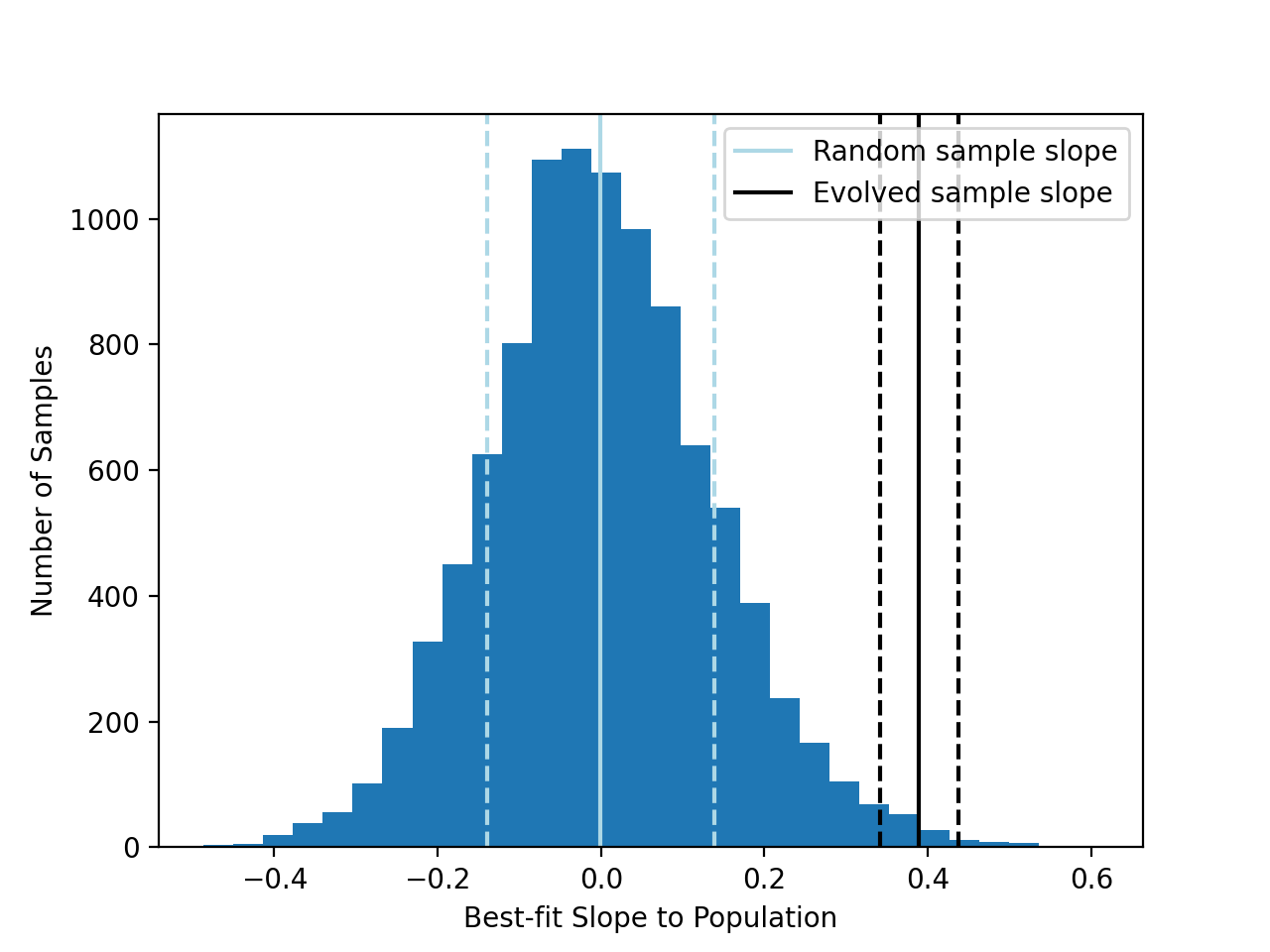}
    \caption{A comparison between the slope determined via a linear regression to the periods and eccentricities of evolved planets shown in Figure \ref{fig:perecc_v1} (black), versus a sample where random eccentricities are assigned to the evolved planets (blue), for all transiting planets with significant eccentricities. The number of samples in each bin is shown on the y-axis. The actual measured linear relation differs from the randomly determined population at a $>$2-$\sigma$ level.
    \label{fig:histogram}}
\end{figure}

\begin{figure*}[ht!]
    \centering
    \includegraphics[width=.49\textwidth]{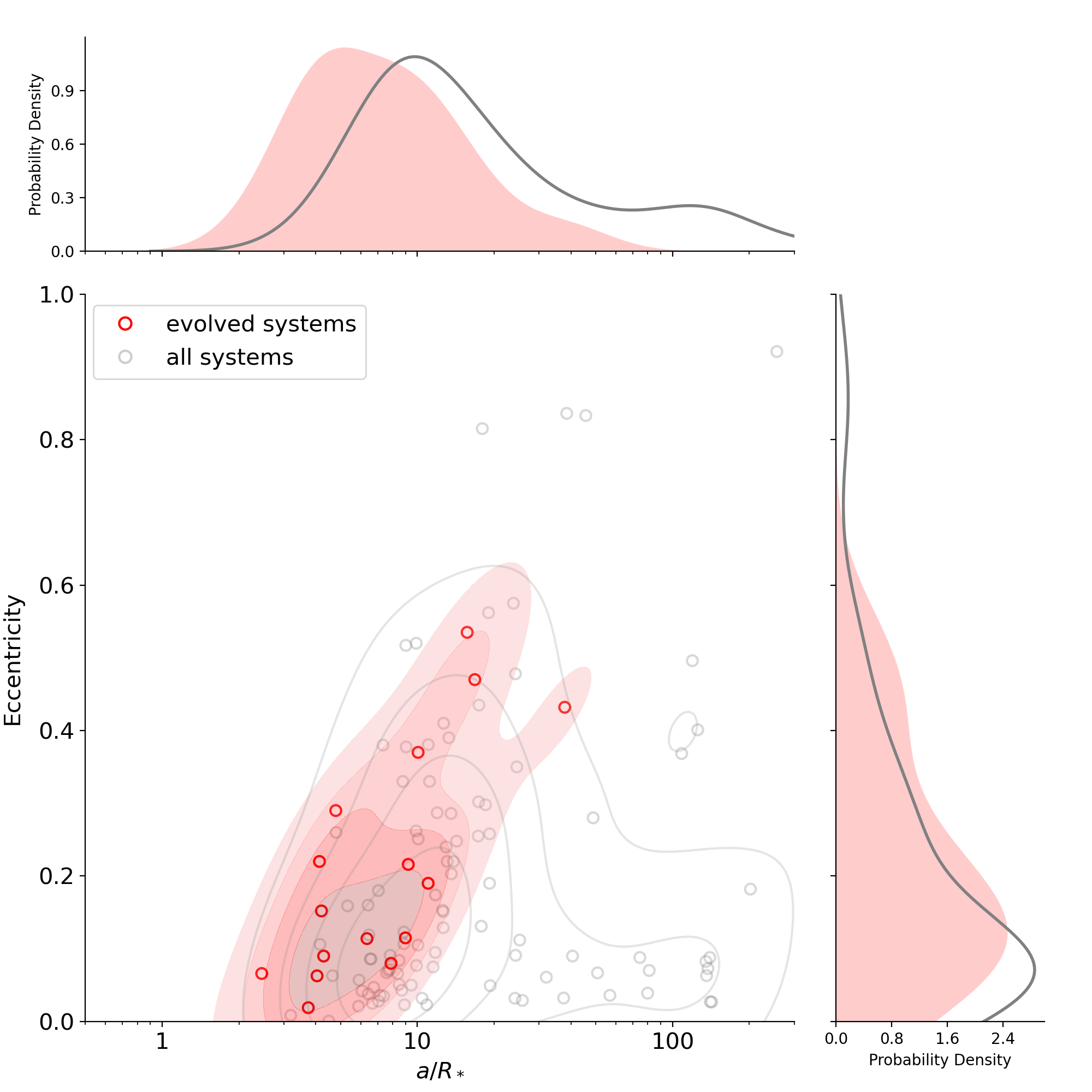}
    \includegraphics[width=.49\textwidth]{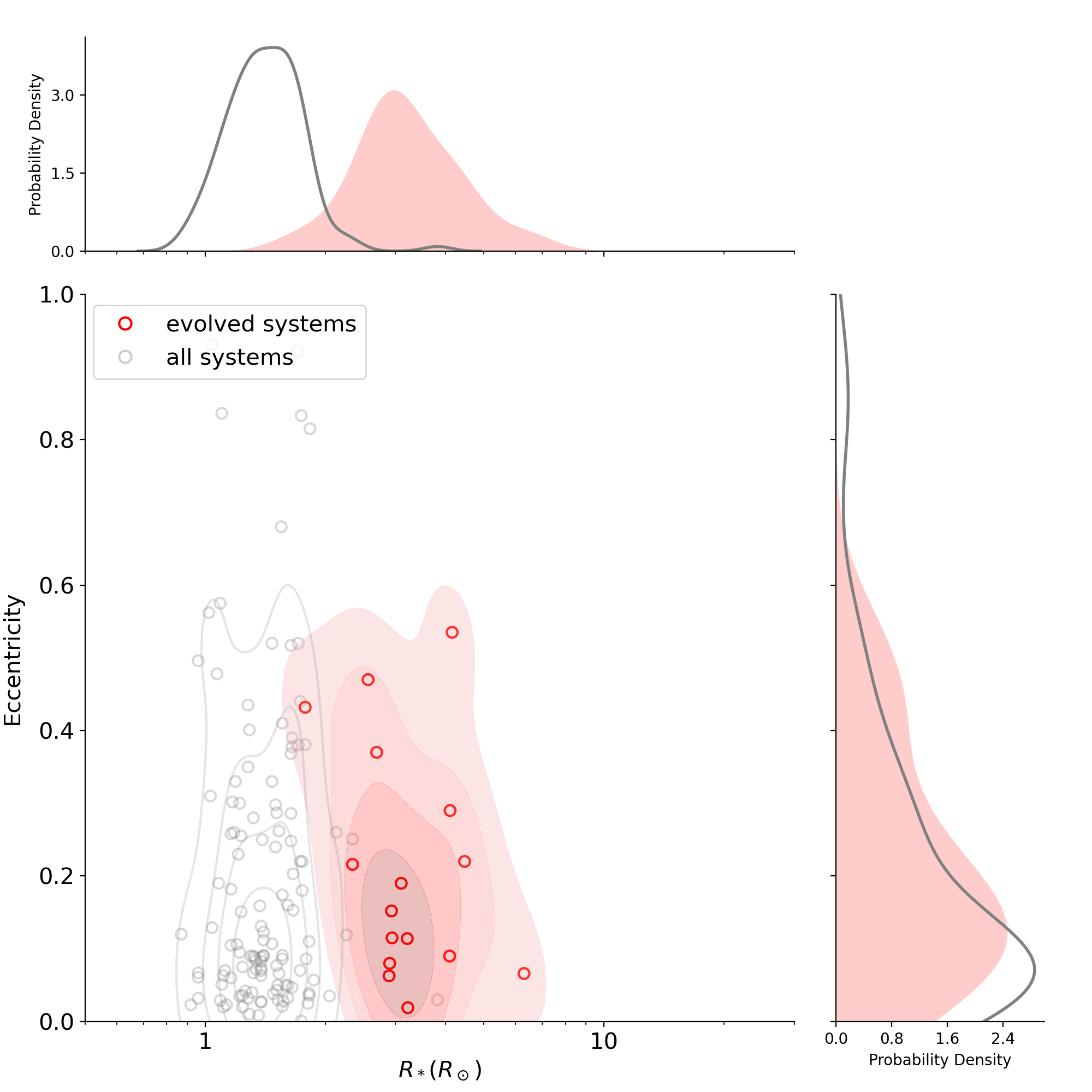}
    \caption{{\it Left:} Orbital eccentricity as a function of semimajor axis divided by stellar radius, highlighting the evolved planet population in red. The evolved population seems to occupy similar ranges in eccentricity and $a$/R$_*$ as the larger total planet population. Statistical tests suggest that differences between planet populations are insignificant in this plane. {\it Right:} Orbital eccentricity as a function of stellar radius for the evolved (red) and overall (gray) planet populations. No clear correlation between stellar radius and eccentricity can be seen. 
    \label{fig:arstarecc}}
\end{figure*}


\begin{deluxetable*}{lcccccccc}
\tabletypesize{\scriptsize}
\tablecaption{Known Giant Planets Transiting Evolved Stars With Significant Orbital Eccentricities\label{tbl-systems}}
\tablewidth{\linewidth}
\tablehead{
\colhead{Name} & \colhead{Planet Mass (M$_\mathrm{J}$) } & \colhead{Planet Radius (R$_\mathrm{J}$) } & \colhead{Orbital Period (d)} & \colhead{Eccentricity} & \colhead{Stellar Mass (M$_\odot$)} & \colhead{Stellar Radius. (R$_\odot$) } & \colhead{Metallicity} & \colhead{Source}
}

\startdata
HD 1397 b & 0.41 $\pm$ 0.02 & 1.03 $\pm$ 0.03 & 11.535 $\pm$ 0.001 & 0.25 $\pm$ 0.03 & 1.32 $\pm$ 0.04 & 2.34 $\pm$ 0.05 & 0.29 $\pm$ 0.09 & 1\\
HD 221416 b & 0.19 $\pm$ 0.02 & 0.84 $\pm$ 0.03 & 14.267 $\pm$ 0.004 & 0.115 $\pm$ 0.03 & 1.21 $\pm$ 0.07 & 2.94 $\pm$ 0.06 & -0.08 $\pm$ 0.08 & 2\\
Kepler-91 b & 0.76 $\pm$ 0.13 & 1.30 $\pm$ 0.07 & 6.246669 $\pm$ 0.00008 & 0.02$^{+0.04}_{-0.02}$ & 1.31 $\pm$ 0.10 & 6.30 $\pm$ 0.16 & 0.11 $\pm$ 0.07 &  3\\
Kepler-432 b & 5.41 $\pm$ 0.32 & 1.15 $\pm$ 0.04  & 52.5011 $\pm$ 0.0001 & 0.51 $\pm$ 0.03 & 1.35 $\pm$ 0.10  & 4.16 $\pm$ 0.12  & -0.07 $\pm$ 0.10 & 4\\
Kepler-435 b & 0.84 $\pm$ 0.15 & 1.99 $\pm$ 0.18  & 8.600153 $\pm$ 0.000002 & 0.11 $\pm$ 0.07 & 1.54 $\pm$ 0.09  & 3.21 $\pm$ 0.30  & -0.18 $\pm$ 0.11 & 5\\
Kepler-643 b & 1.01 $\pm$ 0.20 & 0.91 $\pm$ 0.03  & 16.33889 $\pm$ 0.00001 & 0.37 $\pm$ 0.06 & 1.00 $\pm$ 0.07  & 2.52 $\pm$ 0.07  & 0.16 $\pm$ 0.10 &  6\\
Kepler-1658 b & 5.88 $\pm$ 0.48 & 1.07 $\pm$ 0.05  & 3.84937 $\pm$ 0.00001 & 0.0628 $\pm$ 0.018 & 1.05 $\pm$ 0.15  & 3.91 $\pm$ 0.26  & -0.18 $\pm$ 0.10 &  7\\
K2-39 b & 0.125 $\pm$ 0.014 & 0.51 $\pm$ 0.06  & 4.60543 $\pm$ 0.0005 & 0.15 $\pm$ 0.08 & 1.19 $\pm$ 0.08  &  2.93 $\pm$ 0.21  & 0.32 $\pm$ 0.04 & 8\\
K2-97 b & 0.48 $\pm$ 0.07 & 1.3 $\pm$0.1  & 8.4061 $\pm$ 0.0015 & 0.22 $\pm$ 0.08 & 1.16 $\pm$ 0.12 & 4.20 $\pm$ 0.14  & 0.42 $\pm$ 0.08 & 6\\
K2-99 b & 0.97$\pm$0.09  & 1.29 $\pm$ 0.05  & 18.249 $\pm$ 0.001 & 0.19 $\pm$ 0.04 & 1.60$^{+0.14}_{-0.10}$ & 3.1 $\pm$ 0.1 & 0.20 $\pm$ 0.05 & 9\\
K2-132 b & 0.49 $\pm$ 0.06 & 1.3 $\pm$0.1  & 9.1751 $\pm$ 0.0015 & 0.36 $\pm$ 0.06 & 1.08 $\pm$ 0.08  &  3.85 $\pm$ 0.13  & -0.01$\pm$ 0.08 &  6\\
NGTS-20 b & 2.98 $\pm$ 0.16 & 1.07 $\pm$ 0.04  & 54.18915 $\pm$ 0.00015 & 0.432 $\pm$ 0.023 & 1.47 $\pm$ 0.09 & 1.78 $\pm$ 0.05  & 0.15 $\pm$ 0.08 & 10\\
TOI-2184 b & 0.65 $\pm$ 0.16 & 1.02$\pm$ 0.05  & 6.90683 $\pm$ 0.00009 & 0.08 $\pm$ 0.07 & 1.53 $\pm$ 0.12  & 2.90 $\pm$ 0.14  & 0.14 $\pm$ 0.08 & 11\\
TOI-2337 b & 1.60 $\pm$ 0.15 & 0.9 $\pm$ 0.1  & 2.99432 $\pm$ 0.00008 & 0.019 $\pm$ 0.017 & 1.33 $\pm$ 0.12  & 3.22 $\pm$ 0.06  & 0.39 $\pm$ 0.06 &  12\\
TOI-2669 b & 0.61 $\pm$ 0.19 & 1.76 $\pm$ 0.16  & 6.2034 $\pm$ 0.0001 & 0.09 $\pm$ 0.05 & 1.19 $\pm$ 0.16  & 4.10 $\pm$ 0.04  & 0.10$\pm$ 0.06 & 12\\
TOI-4582 b & 0.53 $\pm$ 0.05 & 0.94$^{+0.09}_{-0.12}$ & 31.034 $\pm$ 0.001 & 0.51 $\pm$ 0.05 & 1.34 $\pm$ 0.02  & 2.51 $\pm$ 0.04  & 0.21 $\pm$ 0.06 & this work\\
\enddata
\tablecomments{1. \citet{nielsen2019}, 2. \citet{huber2019}, 3. \citet{barclay2015}, 4. \citet{quinn2015}, 5. \citet{almenara2015}, 6. \citet{grunblatt2018}, 7. \citet{chontos2019},    8. \citet{vaneylen2016}, 9. \citet{smith2017}, 10. \citet{ulmermoll2022}, 11. \citet{saunders2022}, 12. \citet{grunblatt2022}}
\end{deluxetable*}

\section{Eccentricity Analysis}



\subsection{Eccentricity of \planet}

Both the planet transit and radial velocities measured in \hoststar show strong evidence for an eccentric planetary orbit. Using \texttt{exoplanet} to fit eccentricity using the single-planet system prior defined in \citep{vaneylen2019}, we measure an orbital eccentricity of $e=0.51 \pm 0.05$ for \planet. We note that the determined eccentricity is not sensitive to this choice of prior; the median eccentricity and uncertainties determined strongly agree with the presented values if a uniform prior between 0 and 1 is used instead. The eccentricity of $e=0.51 \pm 0.05$ is among the most eccentric orbits ever found for a planet transiting an evolved star, as well as one of the longest orbital periods found for a planet transiting an evolved star. Only one other known transiting planet around an evolved star, Kepler-432, has a longer orbital period, and interestingly displays a comparable orbital eccentricity \citep{quinn2015, ciceri2015}. Placing these systems in context of the known planets transiting evolved stars suggests that evolved stars may display a correlation between period and eccentricity that does not exist, or at least is not seen as clearly in main sequence systems. 

This may partially be due to the difficulty in detection of long period ($\gtrsim$10 d) planets with \emph{TESS}, as well as the enhanced transit probability of planets on eccentric orbits relative to planets on circular orbits with the same period \citep{nelson1972, barnes2007, beatty2010, barclay2018}. 

However, the boost in transit probability due to eccentricity is relatively modest for all of the planets transiting evolved stars in our sample (a factor of 1.1 at an $e$ = 0.3, or 1.33 at $e$ = 0.5). Looking at orbital periods between 10 and 100 days, two-thirds of the population of planets in our sample have eccentricities larger than 0.2, while only one-third of planets in similar systems transiting similar main sequence stars do. A increase in transit probability of 1.3, larger than the true enhancement for this eccentric evolved population, to all the evolved systems observed is still insufficient to explain this overabundance of eccentric planets. We suggest that this may imply a different eccentricity distribution for planets transiting evolved stars than those transiting main sequence stars, and attempt to test this hypothesis below.

\subsection{Eccentricity of the Evolved Planet Population: Distinction from Main Sequence Systems}

The analysis of orbital eccentricities of exoplanet systems is essential to understanding planetary system formation and evolution \citep[e.g.][]{chatterjee2008, dawson2018}. Earlier studies have shown evidence for dichotomies in the distribution of planet eccentricities between transiting and non-transiting planet populations, single-planet and multiplanet systems, and main sequence and evolved systems  \citep{xie2016, grunblatt2018, vaneylen2019}. This may be driven by stellar evolution, which is known to directly sculpt the planet population through tidal effects, particularly at short periods \citep{hut1981, villaver2014,hamer2019,yee2020}.

However, the effect of stellar evolution on planetary dynamics at longer periods is more unclear, due largely to the small population of known evolved transiting planetary systems. Evolved planet host stars tend to be more massive and metal-rich than most currently known main sequence planet hosts. Enhancement in mass and metallicity might be associated with scattering events during system formation, which can result in a wider range of eccentricities for longer period planets immediately after initial planet formation \citep{dawson2013,frelikh2019}. In addition, other dynamical interactions between planets such as secular angular momentum exchanges and Kozai-Lidov interactions can also excite planet eccentricities after initial formation, and potentially produce hot Jupiters through high-eccentricity migration \citep{petrovich2015a,petrovich2015b, naoz2016}. This process may produce more eccentric, long-period planets during main sequence evolution, resulting in more eccentric planets and hot Jupiters in older systems. Thus, identifying whether the similarities between the main sequence single-planet transiting systems and evolved systems result from differences in intrinsic stellar properties, ages and detection biases, or if the underlying planet populations of main sequence and evolved stars are significantly different, will reveal the more general role of stellar evolution in sculpting planetary architectures. 


Here, we attempt to determine whether these populations are statistically distinct by comparing the transiting planet populations of evolved and main sequence stars, introducing cuts to account for detection biases of planets that transit evolved stars, as well as intrinsic stellar property differences between these two planet populations, and then testing whether the dichotomy remains. We compare the population of planets orbiting evolved (i.e., log($g$) $<$ 3.8) stars as listed in Table \ref{tbl-systems} to the known population of planets (R$_p$ $\geq$ 0.4 R$_\mathrm{J}$) orbiting similar main sequence stars (M$_*$ $\geq$ 1.05 M$_\odot$, [Fe/H] $>$ -0.2 dex) in Figure \ref{fig:perecc_v1}. We show the populations as contours determined from kernel density estimation in the upper panel, as well as the individual planets in each population, where planets transiting evolved stars are shown in red and the larger, overall planet population is shown in gray. We also illustrate the approximate limit for eccentricity growth from planet-planet scattering events as a dashed black line. 

These scattering events can be described as the transfer of the difference in angular velocity of two planets in the same system into angular momentum deficit, where the limiting scattering eccentricity $e_\mathrm{scatter, lim}$ is defined here following the prescription of \citet{dawson2018} as

\begin{equation}
    e_\mathrm{scatter, lim} \approx 0.2 \Big(\frac{M_p}{0.5 M_\mathrm{J}}\Big)^{1/2}\Big(\frac{2 R_\mathrm{J}}{R_p}\Big)^{1/2}\Big(\frac{P}{3 \mathrm{ d}}\Big)^{1/3},
\end{equation}

assuming a typical planet mass of 0.5 M$_\mathrm{J}$ and planet radius of 1.3 R$_\mathrm{J}$. We show an approximate range for high eccentricity migration following the formulation of \citet{dawson2018}, where 

\begin{equation}
a_f = a(1-e^2)
\end{equation}

where $e$ is the initial planet eccentricity, and $a_f$ is the final orbital separation, where we have assumed values of $a_f$ between 0.03 and 0.09 AU, as high-eccentricity migration would take less than 1 Myr for a typical gas giant planet to reach 0.03 AU, yet would take roughly 10 Gyr for the same planet to reach 0.09 AU \citep{dawson2018}.

\begin{figure*}[ht!]
    \centering
    \includegraphics[width=.49\textwidth]{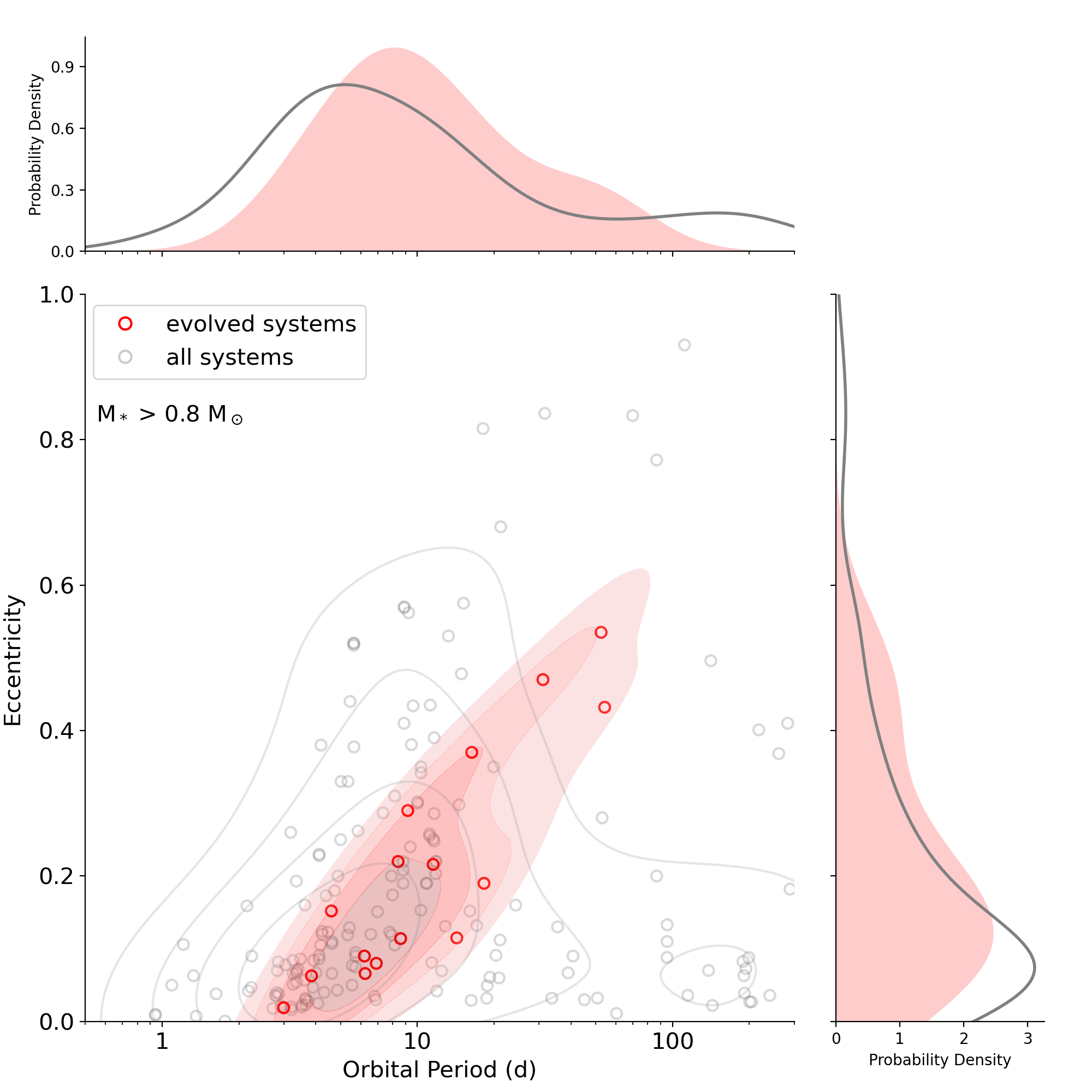}
    \includegraphics[width=.49\textwidth]{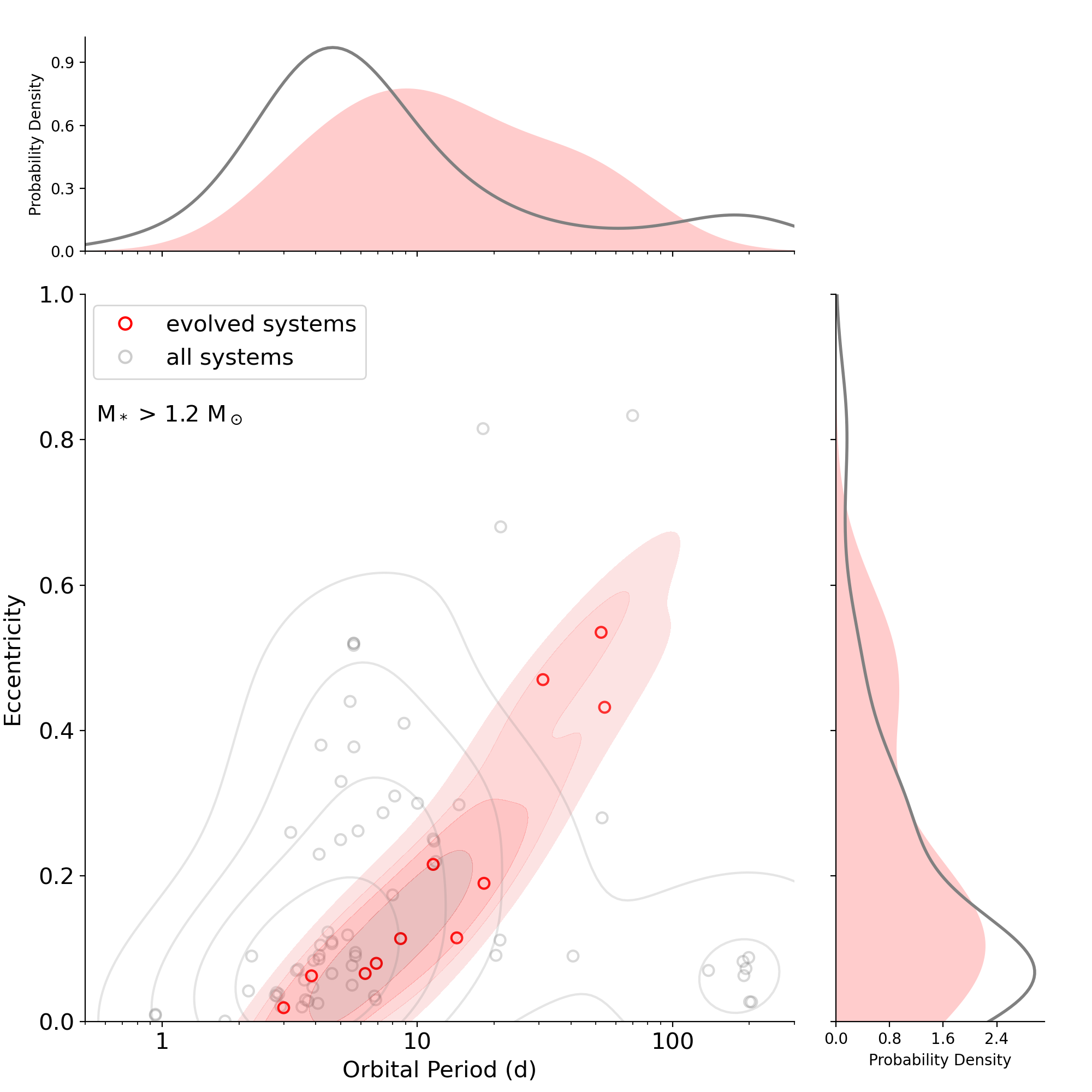}
    \caption{Same as Figure \ref{fig:perecc_v1}, except the cut on stellar mass has been changed, as stellar mass is known to be correlated with planet properties. We use a lower bound of 0.8 M$_\odot$ on the left and 1.2 M$_\odot$ on the right. The difference in planetary architectures between evolved and main sequence systems appears robust with respect to stellar mass.
    \label{fig:perecc_mass}}
\end{figure*}



\begin{figure*}[ht!]
    \centering
    \includegraphics[width=.49\textwidth]{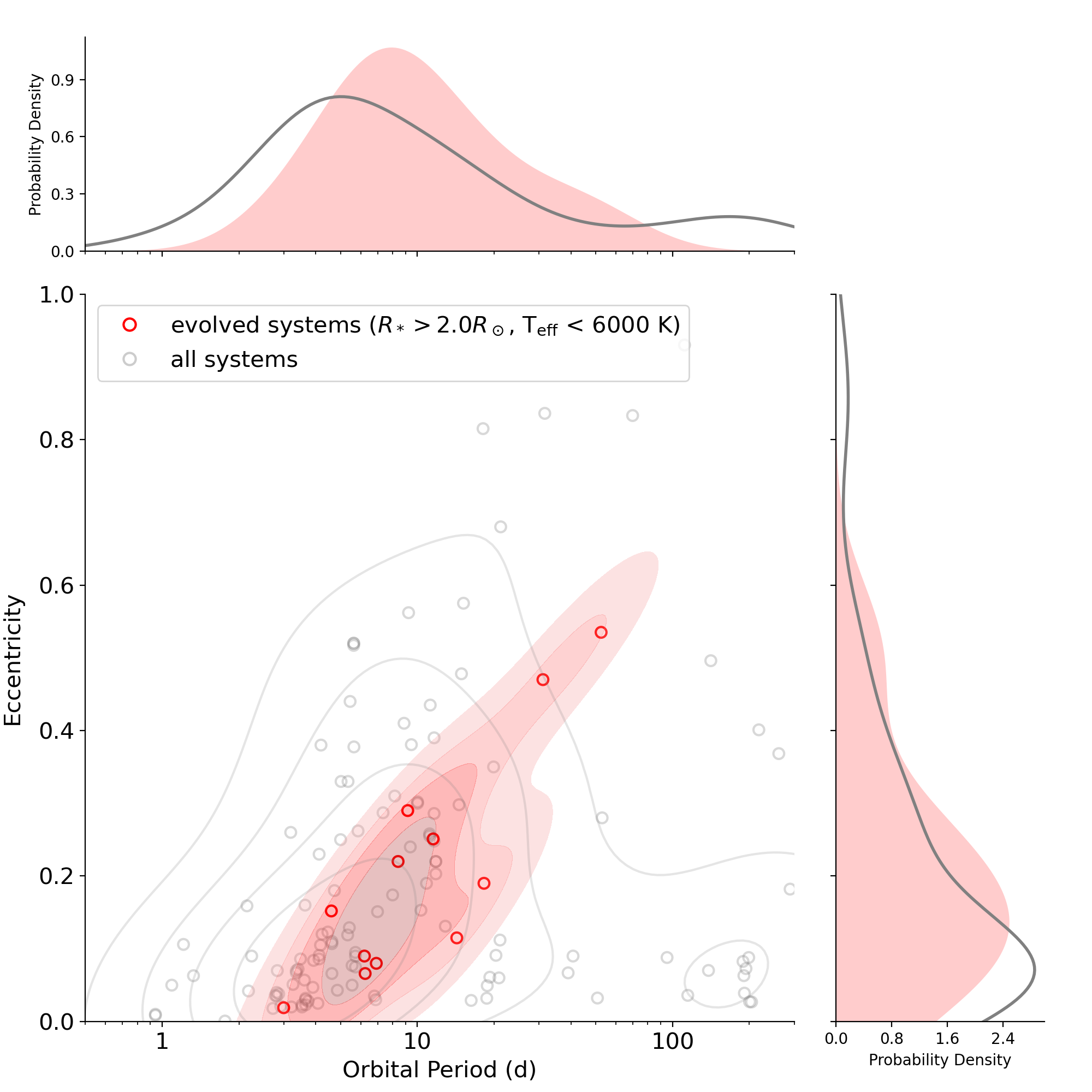}
    \includegraphics[width=.49\textwidth]{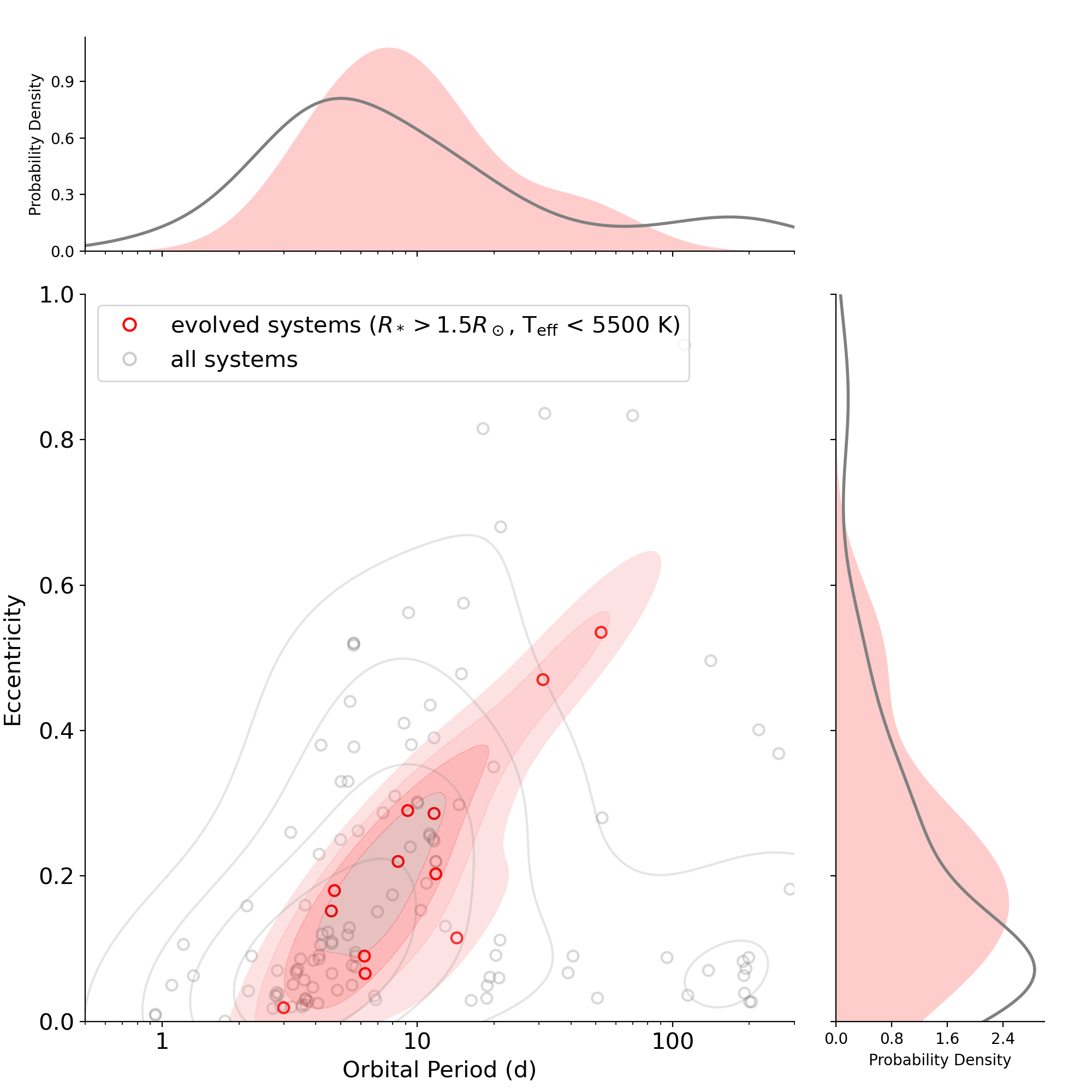}
    \caption{Same as Figure \ref{fig:perecc_v1}, except the population of evolved planets has been altered to instead include systems with host stars with radii $>$ 2.0 R$_\odot$ and effective temperatures T$_\mathrm{eff}$ $<$ 6000 K (left), or systems with host stars with radii $>$ 1.5 R$_\odot$ and effective temperatures T$_\mathrm{eff}$ $<$ 5500 K (right). The period-eccentricity relation seen in evolved systems appears comparably well-constrained using these alternative definitions.
    \label{fig:perecc_evoldefn}}
\end{figure*}


To determine whether these two populations are distinct, we perform a 2-dimensional Kolmogorov-Smirnov test \citep{peacock1983, fasano1987, press2007} between these populations of planets transiting main sequence and evolved stars. We find that the distribution of the population of evolved stars in the semimajor axis-eccentricity plane differs from the population of comparable main sequence (log $g$ $>$ 3.8) planets with a $p$-value of $<$ 0.02, implying the populations are significantly distinct. We also compare the populations of masses and metallicities of the main sequence and evolved host stars studied here, and find that both comparisons have a $p$-value $>$ 0.20, implying that the masses and metallicities of the stellar host populations are not drawn from significantly different distributions.

In addition, we fit a linear regression to the period-eccentricity relation of the planet population orbiting evolved stars. We determine that the eccentricity of a planet in the evolved population $e \propto 0.38\times\mathrm{log}_{10}(P/days) - 0.18$ for orbital periods between 3 and 1000 days. We find a Pearson correlation coefficient $r$=0.88 for this population, and a standard deviation from the estimated slope of 0.05.  We also determine a similar fit to the larger planet population, and find a relation where $e \propto 0.12\times\mathrm{log}_{10}(P/days)$, with a Pearson correlation coefficient $r$=0.35, and a standard deviation from the estimated slope of 0.033, indicating that the period-eccentricity trends seen for single-planet transiting evolved systems differs from that of all similar mass and metallicity systems at $>$5-$\sigma$ significance. 

To test whether the observed features of this population are due to random chance, we produce a random permutation of the set of observed giant planet eccentricities, and test how often the slope we recover for the observed evolved population is found in this reshuffled evolved population. We find that after 10000 random resortings of eccentricities, we recover a best-fit slope of 0.0 $\pm$ 0.15, implying that our recovered eccentricity relation for evolved stars disagrees with a random population at a $>$ 2-$\sigma$ level. We illustrate the results of our 10000 fits to random draws in eccentricity in Figure \ref{fig:histogram}.

Furthermore, we consider whether the observed trend for evolved systems could be due to detection biases. Planet detection via the transit method around evolved stars is biased toward detecting planets that transit more often, and have larger transit depths. In addition, as specified in \citet{saunders2022}, we smooth our light curves which are searched for transits with a median filter with a width of 2 days, which prevents the detection of transits that are longer than 48 hours. These effects will bias us toward finding shorter period and larger planets around smaller stars, and thus the non-detection of short period, eccentric planets around evolved stars cannot be due to a detection bias. As the length of a transit of a planet on an eccentric orbit depends on the argument of periastron $\omega$ (i.e., whether the transit occurs while the planet is at periastron, apastron or in between), and transit duration T$_0$ = T$_\mathrm{circ}$ $\times$ $\sqrt{1-e^2}$/($1+e \mathrm{cos} \omega$), where T$_\mathrm{circ}$ is the transit duration of a planet on an equivalent circular orbits, the change in transit duration for any value of $e \lesssim 0.5$ is less than a factor of 2. Furthermore, eccentric orbits result in a high transit probability with a ``boost factor'' proportional to 1/(1-$e^2$), boosting transit probabilities by a factor of 1.33 or less for an eccentricity of 0.5 or less \citep{barnes2007, beatty2010}. These relatively small variations in transit probability and duration at all eccentricities observed in our sample implies that transit detection is largely independent of eccentricity if $e < 0.5$ at a given orbital period, and thus our non-detection of planets on circular orbits at periods beyond 10 days where we do detect eccentric planets is also not due to detection bias. 


In Figure \ref{fig:arstarecc}, we illustrate orbital eccentricity as a function of semimajor axis divided by stellar radius, again highlighting the evolved planet population. We find that the evolved planet population is largely indistinguishable from the larger planet population in this plane, where the only difference in the statistical distribution can be seen at the largest $a/R_*$ values, which corresponds to periods that are too large to be detectable around many evolved stars. This indicates that the difference in planet orbital period and eccentricity distribution is likely sculpted by processes tied to $a/R_*$ values, such as high-eccentricity migration and tidal interaction. This explains why the shortest period planets around evolved stars are at longer orbits than the shortest period planets around main sequence stars (if $a/R_*$ governs inner boundary for allowed planet orbits, larger stars have a larger minimal separation), but cannot explain the high eccentricities of evolved systems at longer periods.  We find that if we compare the population of planets with $a/R_*$ values $>$ 8, the mass and metallicity distributions overlap with 2-sample K-S test p-values $>$ 0.5, while the 2-dimensional K-S test indicates that in $e$-$a/R_*$ space, the probability of these systems coming from the same distribution $\approx$0.03. Furthermore, we illustrate stellar radius against planet eccentricity in the right-hand panel of this Figure. We see no clear correlation between eccentricity and stellar radius alone, again indicating that stellar radius alone is not driving the observed trend between period and eccentricity for this population, but rather that the trend at short periods is driven by tidal interactions governed by the $a/R_*$ relation.

In general, these trends seen with stellar radius and age follow predicted planet orbital evolution theory: planet circularization and inspiral at small $a$/R$_*$ is most strongly affected by stellar evolution due to exchange of angular momentum between the planet and the star \citep{villaver2014}, while planet-planet scattering events are driven by secular chaos and thus become more likely over time \citep{veras2013}. However, planet-planet scattering can also be impacted by the mass and composition of the host star and planets involved--planets formed around more massive and metal-rich stars tend  to have higher eccentricities \citep{frelikh2019}.  We investigate this in more depth in Figure \ref{fig:perecc_mass}, where we adjust our stellar mass cuts to 0.8 M$_\odot$ (left) and 1.2 M$_\odot$ (right). 

We find that the population of evolved and non-evolved massive stars overlap in $e$-$a$/R$_*$ space and cannot be statistically distinguished. This suggests that stellar mass differences may also contribute to the observed period-eccentricity trend for evolved systems seen here, but does not explain the relatively high eccentricities of planets orbiting somewhat lower-mass (1.0-1.2 M$_\odot$) evolved stars in our sample. We repeat similar tests with metallicity cuts, replacing our previous cut with metallicity values of 0.0 and 0.2. Qualitatively, these cuts result in even tighter linear regressions (with Pearson $r$-values $>$ 0.95), but given the small samples of evolved stars considered ($\lesssim$ 5), the statistical significance of this population difference is low. Thus we do not show these relations in this manuscript, but encourage future studies of this relation once more metal-rich evolved stars hosting transiting planets have been confirmed.

We then redefine the population of `evolved' systems by stellar effective temperature and radius in Figure \ref{fig:perecc_evoldefn}. We define stars as evolved if R$_*$ $>$ 2.0 R$_\odot$ and T$_\mathrm{eff}$ $<$ 6000 K on the left, and R$_*$ $>$ 1.5 R$_\odot$ and T$_\mathrm{eff}$ $<$ 5500 K on the right. The redefinition of evolved to R$_*$ $>$ 1.5 R$_\odot$ and T$_\mathrm{eff}$ $<$ 5500 K  results in the inclusion of HD 89345 b, K2-108 b, and K2-261 b along with a subset of stars from Table 4 \citep{vaneylen2018,petigura2018,ikwutukwa2020}, whereas defining evolved as R$_*$ $>$ 2.0 R$_\odot$ and T$_\mathrm{eff}$ $<$ 6000 K only results in the exclusion of some stars in Table 4. We find that overall, both of these definitions of evolved stars appear to match well with our earlier definition of evolved stars, where the contour morphology is very similar, implying that our selection of evolved stars, as well as this observed dichotomy between main sequence and the evolved stars population, is robust.

We also note that we have restricted our comparisons to planets with radii larger than 0.4 R$_\mathrm{J}$ to avoid the issues of low completeness for transit detection of small planets orbiting evolved stars. However, the period-eccentricity distribution of smaller planets orbiting evolved stars may provide additional evidence confirming or refuting the trends seen here. In particular \citet{jofre2020} measured the eccentricities of two small planets in the evolved system Kepler-276, and found that both orbits were highly eccentric, with eccentricities $>$0.6. However, these planets were also found to be interacting with one another via transit timing variations, which suggests their eccentricities may governed by planet-planet interactions, as opposed to planet-star interaction. In contrast, the planets of the evolved system Kepler-56 appear to have eccentricities which are significantly lower than other planets orbiting evolved stars at similar orbital periods \citep{huber2013b, otor2016}.

Furthermore, planets which have been confirmed orbiting evolved stars using only the radial velocity method are suggested to feature low eccentricity orbits, even at periods $>$10 days \citep[e.g.][]{takarada2018, wolthoff2022}. However, these planets tend to be more massive and at wider orbital separations, and orbit more evolved stars than the transiting planet sample. In addition, the detection biases for radial velocity planet detection and orbital eccentricity characterization is much more difficult to disentangle than for transiting planets due to the uneven time sampling of radial velocity observations, and thus comparing the two planet populations in the period-eccentricity plane is not straightforward. The actual existence of up to 50\% of the RV-only detected planets orbiting evolved stars has recently been questioned due to similar periods seen in magnetic activity indicators within the stellar spectra \citep{delgadomena2018}. Thus, we exclude planets without a transit measurement from our period-eccentricity analysis performed here, and highlight the importance of a search for transiting planets on longer periods around evolved stars to properly characterize the orbital evolution of planets with separations beyond 0.5 AU. 

Theory has suggested that around an orbital distance of $\sim$1 AU, a unstable equilibrium point is reached for the motion of planets with respect to stellar evolution, where planets within 1 AU migrate inward and must eventually be engulfed by their host star, while planets beyond 1 AU migrate outward due to stellar mass loss and can survive stellar evolution through the helium flash, as their orbits are always beyond the stellar surface and tidal forces never become particularly strong for this population \citep{zahn1977, villaver2014}. A search for planets on these periods around evolved stars can reveal whether stellar evolution creates a dearth of planets at these orbits, or whether planets on these orbits are perhaps more stable, or stable for a longer period of time than theories predict.

\section{Orbital Decay of the Evolved Planet Population}\label{sec:orbdec}


The expected tidal interaction between hot gas giant planets and evolved host stars is expected to result in rapid orbital decay and eventual engulfment of the planet. However, orbital decay has only been directly measured in one planetary system to date, which is significantly less evolved than the systems studied here \citep{yee2020,turner2021,wong2022}. By constraining the rate of orbital decay in these systems, we can measure the strength of star-planet tidal interactions and their dependence on star and planet properties.


\begin{figure}[ht!]
    \centering
    \includegraphics[width=.475\textwidth]{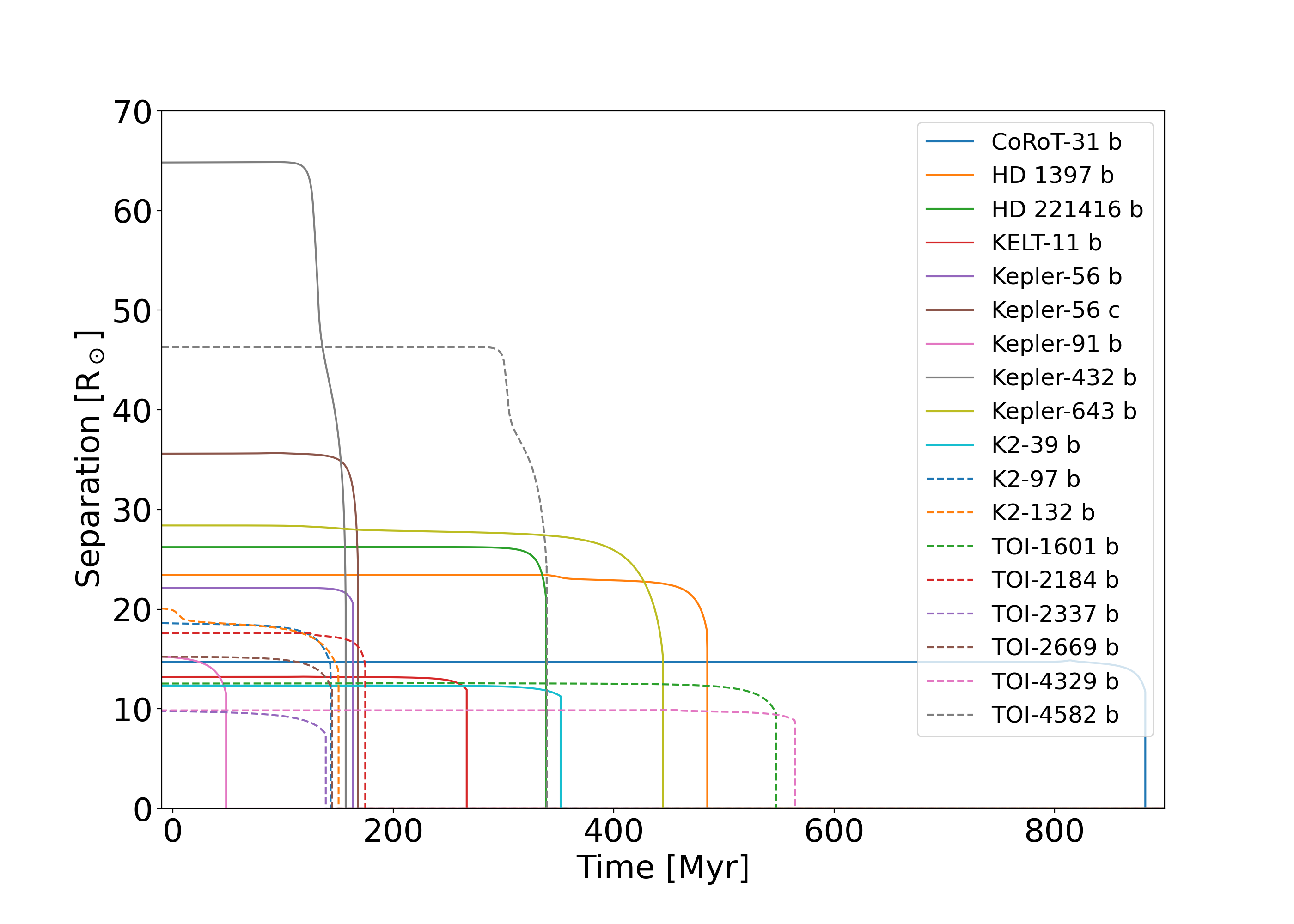}
    \caption{Expected orbital evolution of all evolved planetary systems considered in this study. Time into the future is shown on the x-axis, while orbital separation is shown on the y-axis. These systems show a wide range of orbital decay timescales spanning over an order of magnitude, but do not appear to be decaying in a clear sequence according to period. This implies that the period-eccentricity relation identified is not an active evolutionary path for planetary systems undergoing circularization and inspiral, as there is no clear correlation between orbital period and inspiral timescale.}
    \label{fig:cosmicsep_now}
\end{figure}


Figure \ref{fig:cosmicsep_now} illustrates the expected orbital evolution of the evolved systems studied here over time as a function of orbital separation, calculated using the COSMIC binary evolution code package \citep{breivik2020}, a binary population synthesis package based on BSE \citep{hurley2000, hurley2002}. COSMIC includes the effects of the equilibrium tide with convective damping based on the standard approximations from \citet{zahn1977} and \citet{1989A&A...220..112Z} following the formalism of \citet{hut1981} and thus does not provide direct constraints on the static tidal quality factor $Q`_*$ of the host stars, but does account for the evolution of stellar structure over time, which governs the timescale of orbital decay. We find that all of these systems are expected to experience runaway inspiral in less than 1 Gyr, but the range in inspiral timescales for this population is quite large. Based on the results of COSMIC simulations for this sample of systems, planets on longer period orbits do not seem to survive longer than planets on shorter period orbits, indicating that the longer-period systems are not an earlier stage of the shorter-period systems in our sample, and the period-eccentricity relation identified here is not an evolutionary sequence or pathway for these systems. We also note that the evolutionary timescales listed here are longer than what has been predicted by other binary evolution studies \citep[e.g.][]{sun2018}, which is likely due to the treatment of dynamical tides and wave propagation within a star as it evolves along the sub-giant and giant branches of evolution. Stellar winds and outflows and their associated drag forces may also play a role in inspiral at the latest stages \citep{macleod2018}.

\section{Conclusions} \label{sec:conclusions}

In this manuscript, we have reported the discovery and characterization of \planet, the latest planet discovered in our search for planets transiting evolved stars using \tess\ Full Frame Image data. Our main conclusions based on this discovery are listed as follows:

\begin{itemize}
    \item \planet has one of the longest-period and most eccentric orbits found for a planet transiting an evolved star (log($g$) $<$ 3.8).
    \item The planet population transiting evolved stars appears to follow a log-linear trend in the period-eccentricity plane. This location may be explained by star-planet interaction causing tidal circularization and inspiral at small periods and planet-planet scattering events exciting orbital eccentricities at long periods.
    \item Considering systems around evolved stars with one transiting planet, we find these systems appear to follow a period-eccentricity correlation, where $e$ $\approx$ 0.44 log$_{10}(P{\rm (days)})$ - 0.25 with a Pearson $r$ correlation value $>$ 0.85 and a standard deviation of 0.06. This is inconsistent with a similar linear fit to the population of all comparable planets at $>$5-$\sigma$ significance. A random draw of planetary eccentricities using a similar number of planets as included in the evolved population studied here suggests this trend is inconsistent with the null hypothesis at a $>$ 2-$\sigma$ significance. Additional tests using only higher-mass stars as well as slightly different definitions for an `evolved' stellar host suggest that these trends are robust.
    
\end{itemize}

The \planet system highlights the importance of characterizing the longer period transiting planet population of evolved stars to understand planetary system stability, as well as the need for a focused search to find systems with such long duration transits. The extended \tess\ Missions will extend continuous observations of hundreds of thousands of similar targets to 50 days or more of coverage. Furthermore, the higher cadence and longer baseline Extended Mission observations of \planet may allow asteroseismic characterization of this and similar systems \citep{grunblatt2022}. Similar continuous coverage of evolved stars with next generation surveys such as PLATO will allow for the detection of smaller, less massive planets transiting evolved stars at longer periods \citep{rauer2014, veras2015}. Constraining the orbital properties of these longer-period, smaller planets will be essential for predicting the stability of rocky planets like our own around evolving stars, as well as around more massive stars. Observation of these stars will also be possible at greater distances because of their intrinsic brightnesses \citep{malmquist1922}, allowing comparison of planet demographics between Galactic thin disk and thick disk stars, as well as in other Galactic regions, where stellar populations are known to have different intrinsic properties.





\acknowledgements{ We acknowledge the members of the Astro Data Group at the Center for Computational Astrophysics and the Stellar Rotation group at the American Museum of Natural History for very helpful discussions. We acknowledge the use of public TESS data from pipelines at the TESS Science Office and at the TESS Science Processing Operations Center. Resources supporting this work were provided by the NASA High-End Computing (HEC) Program through the NASA Advanced Supercomputing (NAS) Division at Ames Research Center for the production of the SPOC data products. This work was supported by a NASA Keck PI Data Award, administered by the NASA Exoplanet Science Institute. Data presented herein were obtained at the W. M. Keck Observatory from telescope time allocated to the National Aeronautics and Space Administration through the agency's scientific partnership with the California Institute of Technology and the University of California. The Observatory was made possible by the generous financial support of the W. M. Keck Foundation. The authors wish to recognize and acknowledge the very significant cultural role and reverence that the summit of Maunakea has always had within the indigenous Hawaiian community. We are most fortunate to have the opportunity to conduct observations from this mountain. S.G., N.S., and D.H. acknowledge support by the National Aeronautics and Space Administration under Grants 80NSSC19K0593 and 80NSSC21K0781 issued through the TESS Guest Investigator Program. D.H. acknowledges support from the Alfred P. Sloan Foundation and the National Aeronautics and Space Administration (80NSSC21K0652), and the National Science Foundation (80NSSC21K0652). N.S. acknowledges support from the National  Science Foundation through the Graduate Research Fellowship Program under Grants 1842402 and DGE-1752134. D.V. gratefully acknowledges the support of the STFC via an Ernest Rutherford Fellowship (grant ST/P003850/1). B.S.S. and I.A.S. acknowledge the support of Ministry of Science and Higher Educationof the Russian Federation under the grant 075-15-2020-780 (N13.1902.21.0039). D. D. acknowledges support from the TESS Guest Investigator Program grants 80NSSC21K0108 and 80NSSC22K0185. Any opinions, findings, and conclusions or recommendations expressed in this material are those of the authors and do not necessarily reflect the views of the National Science Foundation. This research has made use of the Exoplanet Follow-up Observation Program website, which is operated by the California Institute of Technology, under contract with the National Aeronautics and Space Administration under the Exoplanet Exploration Program. Funding for the TESS mission is provided by NASA's Science Mission Directorate. 
}

\software{This work relied heavily on open source software tools, and we would like to thank the developers for their contributions to the astronomy community. For data access and de-trending, this research made use of \lightkurve, a Python package for \kepler and \tess data analysis \citep{lightkurve}, \tesscut, a MAST tool for extracting observations from \tess FFIs \citep{brasseur2019}, and \giants, a pipeline for producing and de-trending \tess FFI light curves \citep{saunders2022}. The analysis portion of this research relied on \astropy \citep{astropy2013,astropy2018}, as well as \exoplanet \citep{exoplanet:exoplanet} and its dependencies \citep{exoplanet:agol19, exoplanet:exoplanet, exoplanet:kipping13, exoplanet:luger18, exoplanet:pymc3, exoplanet:theano}.}

\nocite{tange2018}
\clearpage
\bibliography{references}{}

\begin{thebibliography}{}
\expandafter\ifx\csname natexlab\endcsname\relax\def\natexlab#1{#1}\fi
\providecommand{\url}[1]{\href{#1}{#1}}
\providecommand{\dodoi}[1]{doi:~\href{http://doi.org/#1}{\nolinkurl{#1}}}
\providecommand{\doeprint}[1]{\href{http://ascl.net/#1}{\nolinkurl{http://ascl.net/#1}}}
\providecommand{\doarXiv}[1]{\href{https://arxiv.org/abs/#1}{\nolinkurl{https://arxiv.org/abs/#1}}}

\bibitem[{{Agol} {et~al.}(2019){Agol}, {Luger}, \&
  {Foreman-Mackey}}]{exoplanet:agol19}
{Agol}, E., {Luger}, R., \& {Foreman-Mackey}, D. 2019, arXiv e-prints

\bibitem[{{Almenara} {et~al.}(2015){Almenara}, {Damiani}, {Bouchy}, {Havel},
  {Bruno}, {H{\'e}brard}, {Diaz}, {Deleuil}, {Barros}, {Boisse}, {Bonomo},
  {Montagnier}, \& {Santerne}}]{almenara2015}
{Almenara}, J.~M., {Damiani}, C., {Bouchy}, F., {et~al.} 2015, \aap, 575, A71,
  \dodoi{10.1051/0004-6361/201424291}

\bibitem[{{Astropy Collaboration} {et~al.}(2013){Astropy Collaboration},
  {Robitaille}, {Tollerud}, {Greenfield}, {Droettboom}, {Bray}, {Aldcroft},
  {Davis}, {Ginsburg}, {Price-Whelan}, {Kerzendorf}, {Conley}, {Crighton},
  {Barbary}, {Muna}, {Ferguson}, {Grollier}, {Parikh}, {Nair}, {Unther},
  {Deil}, {Woillez}, {Conseil}, {Kramer}, {Turner}, {Singer}, {Fox}, {Weaver},
  {Zabalza}, {Edwards}, {Azalee Bostroem}, {Burke}, {Casey}, {Crawford},
  {Dencheva}, {Ely}, {Jenness}, {Labrie}, {Lim}, {Pierfederici}, {Pontzen},
  {Ptak}, {Refsdal}, {Servillat}, \& {Streicher}}]{astropy2013}
{Astropy Collaboration}, {Robitaille}, T.~P., {Tollerud}, E.~J., {et~al.} 2013,
  \\aap, 558, A33, \dodoi{10.1051/0004-6361/201322068}

\bibitem[{{Barclay} {et~al.}(2015){Barclay}, {Endl}, {Huber}, {Foreman-Mackey},
  {Cochran}, {MacQueen}, {Rowe}, \& {Quintana}}]{barclay2015}
{Barclay}, T., {Endl}, M., {Huber}, D., {et~al.} 2015, \apj, 800, 46,
  \dodoi{10.1088/0004-637X/800/1/46}

\bibitem[{Barclay {et~al.}(2018)Barclay, Pepper, \& Quintana}]{barclay2018}
Barclay, T., Pepper, J., \& Quintana, E.~V. 2018, The Astrophysical Journal
  Supplement Series, 239, 2, \dodoi{10.3847/1538-4365/aae3e9}

\bibitem[{{Barnes}(2007)}]{barnes2007}
{Barnes}, J.~W. 2007, \pasp, 119, 986, \dodoi{10.1086/522039}

\bibitem[{{Beatty} \& {Seager}(2010)}]{beatty2010}
{Beatty}, T.~G., \& {Seager}, S. 2010, \apj, 712, 1433,
  \dodoi{10.1088/0004-637X/712/2/1433}

\bibitem[{{Bellm} {et~al.}(2019){Bellm}, {Kulkarni}, {Graham}, {Dekany},
  {Smith}, {Riddle}, {Masci}, {Helou}, {Prince}, {Adams}, {Barbarino},
  {Barlow}, {Bauer}, {Beck}, {Belicki}, {Biswas}, {Blagorodnova}, {Bodewits},
  {Bolin}, {Brinnel}, {Brooke}, {Bue}, {Bulla}, {Burruss}, {Cenko}, {Chang},
  {Connolly}, {Coughlin}, {Cromer}, {Cunningham}, {De}, {Delacroix}, {Desai},
  {Duev}, {Eadie}, {Farnham}, {Feeney}, {Feindt}, {Flynn}, {Franckowiak},
  {Frederick}, {Fremling}, {Gal-Yam}, {Gezari}, {Giomi}, {Goldstein},
  {Golkhou}, {Goobar}, {Groom}, {Hacopians}, {Hale}, {Henning}, {Ho}, {Hover},
  {Howell}, {Hung}, {Huppenkothen}, {Imel}, {Ip}, {Ivezi{\'c}}, {Jackson},
  {Jones}, {Juric}, {Kasliwal}, {Kaspi}, {Kaye}, {Kelley}, {Kowalski},
  {Kramer}, {Kupfer}, {Landry}, {Laher}, {Lee}, {Lin}, {Lin}, {Lunnan},
  {Giomi}, {Mahabal}, {Mao}, {Miller}, {Monkewitz}, {Murphy}, {Ngeow},
  {Nordin}, {Nugent}, {Ofek}, {Patterson}, {Penprase}, {Porter}, {Rauch},
  {Rebbapragada}, {Reiley}, {Rigault}, {Rodriguez}, {van Roestel}, {Rusholme},
  {van Santen}, {Schulze}, {Shupe}, {Singer}, {Soumagnac}, {Stein}, {Surace},
  {Sollerman}, {Szkody}, {Taddia}, {Terek}, {Van Sistine}, {van Velzen},
  {Vestrand}, {Walters}, {Ward}, {Ye}, {Yu}, {Yan}, \& {Zolkower}}]{bellm2019}
{Bellm}, E.~C., {Kulkarni}, S.~R., {Graham}, M.~J., {et~al.} 2019, \pasp, 131,
  018002, \dodoi{10.1088/1538-3873/aaecbe}

\bibitem[{{Borucki} {et~al.}(2010){Borucki}, {Koch}, {Basri}, {Batalha},
  {Brown}, {Caldwell}, {Caldwell}, {Christensen-Dalsgaard}, {Cochran},
  {DeVore}, {Dunham}, {Dupree}, {Gautier}, {Geary}, {Gilliland}, {Gould},
  {Howell}, {Jenkins}, {Kondo}, {Latham}, {Marcy}, {Meibom}, {Kjeldsen},
  {Lissauer}, {Monet}, {Morrison}, {Sasselov}, {Tarter}, {Boss}, {Brownlee},
  {Owen}, {Buzasi}, {Charbonneau}, {Doyle}, {Fortney}, {Ford}, {Holman},
  {Seager}, {Steffen}, {Welsh}, {Rowe}, {Anderson}, {Buchhave}, {Ciardi},
  {Walkowicz}, {Sherry}, {Horch}, {Isaacson}, {Everett}, {Fischer}, {Torres},
  {Johnson}, {Endl}, {MacQueen}, {Bryson}, {Dotson}, {Haas}, {Kolodziejczak},
  {Van Cleve}, {Chandrasekaran}, {Twicken}, {Quintana}, {Clarke}, {Allen},
  {Li}, {Wu}, {Tenenbaum}, {Verner}, {Bruhweiler}, {Barnes}, \&
  {Prsa}}]{borucki2010}
{Borucki}, W.~J., {Koch}, D., {Basri}, G., {et~al.} 2010, Science, 327, 977,
  \dodoi{10.1126/science.1185402}

\bibitem[{{Brasseur} {et~al.}(2019){Brasseur}, {Phillip}, {Fleming},
  {Mullally}, \& {White}}]{brasseur2019}
{Brasseur}, C.~E., {Phillip}, C., {Fleming}, S.~W., {Mullally}, S.~E., \&
  {White}, R.~L. 2019, {Astrocut: Tools for creating cutouts of TESS images}.
\newblock \doeprint{1905.007}

\bibitem[{{Breivik} {et~al.}(2020){Breivik}, {Coughlin}, {Zevin}, {Rodriguez},
  {Kremer}, {Ye}, {Andrews}, {Kurkowski}, {Digman}, {Larson}, \&
  {Rasio}}]{breivik2020}
{Breivik}, K., {Coughlin}, S., {Zevin}, M., {et~al.} 2020, \apj, 898, 71,
  \dodoi{10.3847/1538-4357/ab9d85}

\bibitem[{{Brown} {et~al.}(2013){Brown}, {Baliber}, {Bianco}, {Bowman},
  {Burleson}, {Conway}, {Crellin}, {Depagne}, {De Vera}, {Dilday}, {Dragomir},
  {Dubberley}, {Eastman}, {Elphick}, {Falarski}, {Foale}, {Ford}, {Fulton},
  {Garza}, {Gomez}, {Graham}, {Greene}, {Haldeman}, {Hawkins}, {Haworth},
  {Haynes}, {Hidas}, {Hjelstrom}, {Howell}, {Hygelund}, {Lister}, {Lobdill},
  {Martinez}, {Mullins}, {Norbury}, {Parrent}, {Paulson}, {Petry}, {Pickles},
  {Posner}, {Rosing}, {Ross}, {Sand}, {Saunders}, {Shobbrook}, {Shporer},
  {Street}, {Thomas}, {Tsapras}, {Tufts}, {Valenti}, {Vander Horst}, {Walker},
  {White}, \& {Willis}}]{brown2013}
{Brown}, T.~M., {Baliber}, N., {Bianco}, F.~B., {et~al.} 2013, \pasp, 125,
  1031, \dodoi{10.1086/673168}

\bibitem[{{Buchhave} {et~al.}(2012){Buchhave}, {Latham}, {Johansen},
  {Bizzarro}, {Torres}, {Rowe}, {Batalha}, {Borucki}, {Brugamyer}, {Caldwell},
  {Bryson}, {Ciardi}, {Cochran}, {Endl}, {Esquerdo}, {Ford}, {Geary},
  {Gilliland}, {Hansen}, {Isaacson}, {Laird}, {Lucas}, {Marcy}, {Morse},
  {Robertson}, {Shporer}, {Stefanik}, {Still}, \& {Quinn}}]{buchhave2012}
{Buchhave}, L.~A., {Latham}, D.~W., {Johansen}, A., {et~al.} 2012, \nat, 486,
  375, \dodoi{10.1038/nature11121}

\bibitem[{{Chatterjee} {et~al.}(2008){Chatterjee}, {Ford}, {Matsumura}, \&
  {Rasio}}]{chatterjee2008}
{Chatterjee}, S., {Ford}, E.~B., {Matsumura}, S., \& {Rasio}, F.~A. 2008, \apj,
  686, 580, \dodoi{10.1086/590227}

\bibitem[{{Choi} {et~al.}(2016){Choi}, {Dotter}, {Conroy}, {Cantiello},
  {Paxton}, \& {Johnson}}]{choi2016}
{Choi}, J., {Dotter}, A., {Conroy}, C., {et~al.} 2016, ArXiv e-prints.
\newblock \doarXiv{1604.08592}

\bibitem[{{Chontos} {et~al.}(2019){Chontos}, {Huber}, {Latham}, {Bieryla}, {Van
  Eylen}, {Bedding}, {Berger}, {Buchhave}, {Campante}, {Chaplin}, {Colman},
  {Coughlin}, {Davies}, {Hirano}, {Howard}, \& {Isaacson}}]{chontos2019}
{Chontos}, A., {Huber}, D., {Latham}, D.~W., {et~al.} 2019, \aj, 157, 192,
  \dodoi{10.3847/1538-3881/ab0e8e}

\bibitem[{{Ciceri} {et~al.}(2015){Ciceri}, {Lillo-Box}, {Southworth},
  {Mancini}, {Henning}, \& {Barrado}}]{ciceri2015}
{Ciceri}, S., {Lillo-Box}, J., {Southworth}, J., {et~al.} 2015, \aap, 573, L5,
  \dodoi{10.1051/0004-6361/201425145}

\bibitem[{{Dawson} \& {Johnson}(2018)}]{dawson2018}
{Dawson}, R.~I., \& {Johnson}, J.~A. 2018, \araa, 56, 175,
  \dodoi{10.1146/annurev-astro-081817-051853}

\bibitem[{{Dawson} \& {Murray-Clay}(2013)}]{dawson2013}
{Dawson}, R.~I., \& {Murray-Clay}, R.~A. 2013, \apjl, 767, L24,
  \dodoi{10.1088/2041-8205/767/2/L24}

\bibitem[{{Delgado Mena} {et~al.}(2018){Delgado Mena}, {Lovis}, {Santos},
  {Gomes da Silva}, {Mortier}, {Tsantaki}, {Sousa}, {Figueira}, {Cunha},
  {Campante}, {Adibekyan}, {Faria}, \& {Montalto}}]{delgadomena2018}
{Delgado Mena}, E., {Lovis}, C., {Santos}, N.~C., {et~al.} 2018, \aap, 619, A2,
  \dodoi{10.1051/0004-6361/201833152}

\bibitem[{{Dotter}(2016)}]{dotter2016}
{Dotter}, A. 2016, \apjs, 222, 8, \dodoi{10.3847/0067-0049/222/1/8}

\bibitem[{Eisner {et~al.}(2020)Eisner, Barrag{\'a}n, Aigrain, Lintott, Miller,
  Zicher, Boyajian, Brice{\~n}o, Bryant, Christiansen, Feinstein,
  {Flor-Torres}, Fridlund, Gandolfi, Gilbert, Guerrero, Jenkins, Jones,
  Kristiansen, Vanderburg, Law, {L{\'o}pez-S{\'a}nchez}, Mann, Safron, Schwamb,
  Stassun, Osborn, Wang, Zic, Ziegler, Barnet, Bean, Bundy, Chetnik, Dawson,
  Garstone, Stenner, Huten, Larish, Melanson, Mitchell, Moore, Peltsch, Rogers,
  Schuster, Smith, Simister, Tanner, Terentev, \& Tsymbal}]{eisner2020}
Eisner, N.~L., Barrag{\'a}n, O., Aigrain, S., {et~al.} 2020, Monthly Notices of
  the Royal Astronomical Society, staa138, \dodoi{10.1093/mnras/staa138}

\bibitem[{{Fasano} \& {Franceschini}(1987)}]{fasano1987}
{Fasano}, G., \& {Franceschini}, A. 1987, \mnras, 225, 155,
  \dodoi{10.1093/mnras/225.1.155}

\bibitem[{{F{\H{u}}r{\'e}sz} {et~al.}(2008){F{\H{u}}r{\'e}sz}, {Szentgyorgyi},
  \& {Meibom}}]{furesz2008}
{F{\H{u}}r{\'e}sz}, G., {Szentgyorgyi}, A.~H., \& {Meibom}, S. 2008, in
  Precision Spectroscopy in Astrophysics, ed. N.~C. {Santos}, L.~{Pasquini},
  A.~C.~M. {Correia}, \& M.~{Romaniello}, 287--290,
  \dodoi{10.1007/978-3-540-75485-5\_68}

\bibitem[{Foreman-Mackey {et~al.}(2020)Foreman-Mackey, Luger, Czekala, Agol,
  Price-Whelan, \& Barclay}]{exoplanet:exoplanet}
Foreman-Mackey, D., Luger, R., Czekala, I., {et~al.} 2020,
  exoplanet-dev/exoplanet v0.3.2, \dodoi{10.5281/zenodo.1998447}

\bibitem[{{Frelikh} {et~al.}(2019){Frelikh}, {Jang}, {Murray-Clay}, \&
  {Petrovich}}]{frelikh2019}
{Frelikh}, R., {Jang}, H., {Murray-Clay}, R.~A., \& {Petrovich}, C. 2019,
  \apjl, 884, L47, \dodoi{10.3847/2041-8213/ab4a7b}

\bibitem[{{Grunblatt} {et~al.}(2019){Grunblatt}, {Huber}, {Gaidos}, {Hon},
  {Zinn}, \& {Stello}}]{grunblatt2019}
{Grunblatt}, S.~K., {Huber}, D., {Gaidos}, E., {et~al.} 2019, \aj, 158, 227,
  \dodoi{10.3847/1538-3881/ab4c35}

\bibitem[{{Grunblatt} {et~al.}(2016){Grunblatt}, {Huber}, {Gaidos}, {Lopez},
  {Fulton}, {Vanderburg}, {Barclay}, {Fortney}, {Howard}, {Isaacson}, {Mann},
  {Petigura}, {Silva Aguirre}, \& {Sinukoff}}]{grunblatt2016}
{Grunblatt}, S.~K., {Huber}, D., {Gaidos}, E.~J., {et~al.} 2016, \aj, 152, 185,
  \dodoi{10.3847/0004-6256/152/6/185}

\bibitem[{{Grunblatt} {et~al.}(2017){Grunblatt}, {Huber}, {Gaidos}, {Lopez},
  {Howard}, {Isaacson}, {Sinukoff}, {Vanderburg}, {Nofi}, {Yu}, {North},
  {Chaplin}, {Foreman-Mackey}, {Petigura}, {Ansdell}, {Weiss}, {Fulton}, \&
  {Lin}}]{grunblatt2017}
{Grunblatt}, S.~K., {Huber}, D., {Gaidos}, E., {et~al.} 2017, \aj, 154, 254,
  \dodoi{10.3847/1538-3881/aa932d}

\bibitem[{{Grunblatt} {et~al.}(2018){Grunblatt}, {Huber}, {Gaidos}, {Lopez},
  {Barclay}, {Chontos}, {Sinukoff}, {Van Eylen}, {Howard}, \&
  {Isaacson}}]{grunblatt2018}
---. 2018, \apjl, 861, L5, \dodoi{10.3847/2041-8213/aacc67}

\bibitem[{{Grunblatt} {et~al.}(2022){Grunblatt}, {Saunders}, {Sun}, {Chontos},
  {Soares-Furtado}, {Eisner}, {Pereira}, {Komacek}, {Huber}, {Collins}, {Wang},
  {Stockdale}, {Quinn}, {Tronsgaard}, {Zhou}, {Nowak}, {Deeg}, {Ciardi},
  {Boyle}, {Rice}, {Dai}, {Blunt}, {Van Zandt}, {Beard}, {Akana Murphy},
  {Dalba}, {Lubin}, {Polanski}, {Brinkman}, {Howard}, {Buchhave}, {Angus},
  {Ricker}, {Jenkins}, {Wohler}, {Goeke}, {Levine}, {Colon}, {Huang},
  {Kunimoto}, {Shporer}, {Latham}, {Seager}, {Vanderspek}, \&
  {Winn}}]{grunblatt2022}
{Grunblatt}, S.~K., {Saunders}, N., {Sun}, M., {et~al.} 2022, \aj, 163, 120,
  \dodoi{10.3847/1538-3881/ac4972}

\bibitem[{{Guerrero} {et~al.}(2021){Guerrero}, {Seager}, {Huang}, {Vanderburg},
  {Garcia Soto}, {Mireles}, {Hesse}, {Fong}, {Glidden}, {Shporer}, {Latham},
  {Collins}, {Quinn}, {Burt}, {Dragomir}, {Crossfield}, {Vanderspek},
  {Fausnaugh}, {Burke}, {Ricker}, {Daylan}, {Essack}, {G{\"u}nther}, {Osborn},
  {Pepper}, {Rowden}, {Sha}, {Villanueva}, {Yahalomi}, {Yu}, {Ballard},
  {Batalha}, {Berardo}, {Chontos}, {Dittmann}, {Esquerdo}, {Mikal-Evans},
  {Jayaraman}, {Krishnamurthy}, {Louie}, {Mehrle}, {Niraula}, {Rackham},
  {Rodriguez}, {Rowden}, {Sousa-Silva}, {Watanabe}, {Wong}, {Zhan},
  {Zivanovic}, {Christiansen}, {Ciardi}, {Swain}, {Lund}, {Mullally},
  {Fleming}, {Rodriguez}, {Boyd}, {Quintana}, {Barclay}, {Col{\'o}n},
  {Rinehart}, {Schlieder}, {Clampin}, {Jenkins}, {Twicken}, {Caldwell},
  {Coughlin}, {Henze}, {Lissauer}, {Morris}, {Rose}, {Smith}, {Tenenbaum},
  {Ting}, {Wohler}, {Bakos}, {Bean}, {Berta-Thompson}, {Bieryla}, {Bouma},
  {Buchhave}, {Butler}, {Charbonneau}, {Doty}, {Ge}, {Holman}, {Howard},
  {Kaltenegger}, {Kane}, {Kjeldsen}, {Kreidberg}, {Lin}, {Minsky}, {Narita},
  {Paegert}, {P{\'a}l}, {Palle}, {Sasselov}, {Spencer}, {Sozzetti}, {Stassun},
  {Torres}, {Udry}, \& {Winn}}]{guerrero2021}
{Guerrero}, N.~M., {Seager}, S., {Huang}, C.~X., {et~al.} 2021, \apjs, 254, 39,
  \dodoi{10.3847/1538-4365/abefe1}

\bibitem[{{Hamer} \& {Schlaufman}(2019)}]{hamer2019}
{Hamer}, J.~H., \& {Schlaufman}, K.~C. 2019, \aj, 158, 190,
  \dodoi{10.3847/1538-3881/ab3c56}

\bibitem[{{Hattori} {et~al.}(2021){Hattori}, {Foreman-Mackey}, {Hogg},
  {Montet}, {Angus}, {Pritchard}, {Curtis}, \& {Sch{\"o}lkopf}}]{hattori2021}
{Hattori}, S., {Foreman-Mackey}, D., {Hogg}, D.~W., {et~al.} 2021, arXiv
  e-prints, arXiv:2106.15063.
\newblock \doarXiv{2106.15063}

\bibitem[{{Hatzes} {et~al.}(2003){Hatzes}, {Cochran}, {Endl}, {McArthur},
  {Paulson}, {Walker}, {Campbell}, \& {Yang}}]{hatzes2003}
{Hatzes}, A.~P., {Cochran}, W.~D., {Endl}, M., {et~al.} 2003, \apj, 599, 1383,
  \dodoi{10.1086/379281}

\bibitem[{{Howell} {et~al.}(2014){Howell}, {Sobeck}, {Haas}, {Still},
  {Barclay}, {Mullally}, {Troeltzsch}, {Aigrain}, {Bryson}, {Caldwell},
  {Chaplin}, {Cochran}, {Huber}, {Marcy}, {Miglio}, {Najita}, {Smith},
  {Twicken}, \& {Fortney}}]{howell2014}
{Howell}, S.~B., {Sobeck}, C., {Haas}, M., {et~al.} 2014, \pasp, 126, 398,
  \dodoi{10.1086/676406}

\bibitem[{{Huang} {et~al.}(2020){Huang}, {Vanderburg}, {P{\'a}l}, {Sha}, {Yu},
  {Fong}, {Fausnaugh}, {Shporer}, {Guerrero}, {Vanderspek}, \&
  {Ricker}}]{huang2020}
{Huang}, C.~X., {Vanderburg}, A., {P{\'a}l}, A., {et~al.} 2020, Research Notes
  of the American Astronomical Society, 4, 206,
  \dodoi{10.3847/2515-5172/abca2d}

\bibitem[{{Huber} {et~al.}(2013){Huber}, {Carter}, {Barbieri}, {Miglio},
  {Deck}, {Fabrycky}, {Montet}, {Buchhave}, {Chaplin}, {Hekker},
  {Montalb{\'a}n}, {Sanchis-Ojeda}, {Basu}, {Bedding}, {Campante},
  {Christensen-Dalsgaard}, {Elsworth}, {Stello}, {Arentoft}, {Ford},
  {Gilliland}, {Handberg}, {Howard}, {Isaacson}, {Johnson}, {Karoff},
  {Kawaler}, {Kjeldsen}, {Latham}, {Lund}, {Lundkvist}, {Marcy}, {Metcalfe},
  {Silva Aguirre}, \& {Winn}}]{huber2013b}
{Huber}, D., {Carter}, J.~A., {Barbieri}, M., {et~al.} 2013, Science, 342, 331,
  \dodoi{10.1126/science.1242066}

\bibitem[{{Huber} {et~al.}(2017){Huber}, {Zinn}, {Bojsen-Hansen},
  {Pinsonneault}, {Sahlholdt}, {Serenelli}, {Silva Aguirre}, {Stassun},
  {Stello}, {Tayar}, {Bastien}, {Bedding}, {Buchhave}, {Chaplin}, {Davies},
  {Garc{\'{\i}}a}, {Latham}, {Mathur}, {Mosser}, \& {Sharma}}]{huber2017}
{Huber}, D., {Zinn}, J., {Bojsen-Hansen}, M., {et~al.} 2017, \apj, 844, 102,
  \dodoi{10.3847/1538-4357/aa75ca}

\bibitem[{{Huber} {et~al.}(2019){Huber}, {Chaplin}, {Chontos}, {Kjeldsen},
  {Christensen-Dalsgaard}, {Bedding}, {Ball}, {Brahm}, {Espinoza}, {Henning},
  {Jordan}, {Sarkis}, {Knudstrup}, {Albrecht}, {Grundahl}, {Fredslund
  Andersen}, {Palle}, {Crossfield}, {Fulton}, {Howard}, {Isaacson}, {Weiss},
  {Handberg}, {Lund}, {Serenelli}, {Mosumgaard}, {Stokholm}, {Bierlya},
  {Buchhave}, {Latham}, {Quinn}, {Gaidos}, {Hirano}, {Ricker}, {Vanderspek},
  {Seager}, {Jenkins}, {Winn}, {Antia}, {Appourchaux}, {Basu}, {Bell},
  {Benomar}, {Bonanno}, {Buzasi}, {Campante}, {Celik Orhan}, {Corsaro},
  {Cunha}, {Davies}, {Deheuvels}, {Grunblatt}, {Hasanzadeh}, {Di Mauro},
  {Garcia}, {Gaulme}, {Girardi}, {Guzik}, {Hon}, {Jiang}, {Kallinger},
  {Kawaler}, {Kuszlewicz}, {Lebreton}, {Li}, {Lucas}, {Lundkvist}, {Mathis},
  {Mathur}, {Mazumdar}, {Metcalfe}, {Miglio}, {Monteiro}, {Mosser}, {Noll},
  {Nsamba}, {Mann}, {Ong}, {Ortel}, {Pereira}, {Ranadive}, {Regulo},
  {Rodrigues}, {Roxburgh}, {Silva Aguirre}, {Smalley}, {Schofield}, {Sousa},
  {Stassun}, {Stello}, {Tayar}, {White}, {Verma}, {Vrard}, {Yildiz}, {Baker},
  {Bazot}, {Beichmann}, {Bergmann}, {Bugnet}, {Cale}, {Carlino}, {Cartwright},
  {Christiansen}, {Ciardi}, {Creevey}, {Dittmann}, {Dias Do Nascimento}, {van
  Eylen}, {Furesz}, {Gagne}, {Gao}, {Gazeas}, {Giddens}, {Hall}, {Hekker},
  {Ireland}, {Latouf}, {LeBrun}, {Levine}, {Matzko}, {Natinsky}, {Page},
  {Plavchan}, {Mansouri-Samani}, {McCauliff}, {Mullally}, {Orenstein}, {Soto},
  {Paegert}, {van Saders}, {Schnaible}, {Soderblom}, {Szabo}, {Tanner},
  {Tinney}, {Teske}, {Thomas}, {Trampedach}, {Wright}, \&
  {Zohrabi}}]{huber2019}
{Huber}, D., {Chaplin}, W.~J., {Chontos}, A., {et~al.} 2019, arXiv e-prints.
\newblock \doarXiv{1901.01643}

\bibitem[{{Hurley} {et~al.}(2000){Hurley}, {Pols}, \& {Tout}}]{hurley2000}
{Hurley}, J.~R., {Pols}, O.~R., \& {Tout}, C.~A. 2000, \mnras, 315, 543,
  \dodoi{10.1046/j.1365-8711.2000.03426.x}

\bibitem[{{Hurley} {et~al.}(2002){Hurley}, {Tout}, \& {Pols}}]{hurley2002}
{Hurley}, J.~R., {Tout}, C.~A., \& {Pols}, O.~R. 2002, \mnras, 329, 897,
  \dodoi{10.1046/j.1365-8711.2002.05038.x}

\bibitem[{{Hut}(1981)}]{hut1981}
{Hut}, P. 1981, \aap, 99, 126

\bibitem[{{Ikwut-Ukwa} {et~al.}(2020){Ikwut-Ukwa}, {Rodriguez}, {Bieryla},
  {Vanderburg}, {Mocnik}, {Kane}, {Quinn}, {Col{\'o}n}, {Zhou}, {Eastman},
  {Huang}, {Latham}, {Dotson}, {Jenkins}, {Ricker}, {Seager}, {Vanderspek},
  {Winn}, {Barclay}, {Barentsen}, {Berta-Thompson}, {Charbonneau}, {Dragomir},
  {Daylan}, {G{\"u}nther}, {Hedges}, {Henze}, {McDermott}, {Schlieder},
  {Quintana}, {Smith}, {Twicken}, \& {Yahalomi}}]{ikwutukwa2020}
{Ikwut-Ukwa}, M., {Rodriguez}, J.~E., {Bieryla}, A., {et~al.} 2020, \aj, 160,
  209, \dodoi{10.3847/1538-3881/aba964}

\bibitem[{{Jenkins}(2002)}]{jenkins2002}
{Jenkins}, J.~M. 2002, \apj, 575, 493, \dodoi{10.1086/341136}

\bibitem[{{Jenkins} {et~al.}(2020){Jenkins}, {Tenenbaum}, {Seader}, {Burke},
  {McCauliff}, {Smith}, {Twicken}, \& {Chandrasekaran}}]{jenkins2020}
{Jenkins}, J.~M., {Tenenbaum}, P., {Seader}, S., {et~al.} 2020, {Kepler Data
  Processing Handbook: Transiting Planet Search}, Kepler Science Document
  KSCI-19081-003

\bibitem[{{Jenkins} {et~al.}(2010){Jenkins}, {Caldwell}, {Chandrasekaran},
  {Twicken}, {Bryson}, {Quintana}, {Clarke}, {Li}, {Allen}, {Tenenbaum}, {Wu},
  {Klaus}, {Middour}, {Cote}, {McCauliff}, {Girouard}, {Gunter}, {Wohler},
  {Sommers}, {Hall}, {Uddin}, {Wu}, {Bhavsar}, {Van Cleve}, {Pletcher},
  {Dotson}, {Haas}, {Gilliland}, {Koch}, \& {Borucki}}]{jenkins2010}
{Jenkins}, J.~M., {Caldwell}, D.~A., {Chandrasekaran}, H., {et~al.} 2010,
  \apjl, 713, L87, \dodoi{10.1088/2041-8205/713/2/L87}

\bibitem[{{Jenkins} {et~al.}(2016){Jenkins}, {Twicken}, {McCauliff},
  {Campbell}, {Sanderfer}, {Lung}, {Mansouri-Samani}, {Girouard}, {Tenenbaum},
  {Klaus}, {Smith}, {Caldwell}, {Chacon}, {Henze}, {Heiges}, {Latham},
  {Morgan}, {Swade}, {Rinehart}, \& {Vanderspek}}]{jenkins2016}
{Jenkins}, J.~M., {Twicken}, J.~D., {McCauliff}, S., {et~al.} 2016, in
  \procspie, Vol. 9913, Software and Cyberinfrastructure for Astronomy IV,
  99133E, \dodoi{10.1117/12.2233418}

\bibitem[{{Jofr{\'e}} {et~al.}(2020){Jofr{\'e}}, {Almenara}, {Petrucci},
  {D{\'\i}az}, {G{\'o}mez Maqueo Chew}, {Martioli}, {Ram{\'\i}rez},
  {Garc{\'\i}a}, {Saffe}, {Canul}, {Buccino}, {G{\'o}mez}, \& {Moreno
  Hilario}}]{jofre2020}
{Jofr{\'e}}, E., {Almenara}, J.~M., {Petrucci}, R., {et~al.} 2020, \aap, 634,
  A29, \dodoi{10.1051/0004-6361/201936446}

\bibitem[{{Jones} {et~al.}(2014){Jones}, {Jenkins}, {Bluhm}, {Rojo}, \&
  {Melo}}]{jones2014}
{Jones}, M.~I., {Jenkins}, J.~S., {Bluhm}, P., {Rojo}, P., \& {Melo}, C.~H.~F.
  2014, \aap, 566, A113, \dodoi{10.1051/0004-6361/201323345}

\bibitem[{{Jones} {et~al.}(2018){Jones}, {Brahm}, {Espinoza}, {Jord{\'a}n},
  {Rojas}, {Rabus}, {Drass}, {Zapata}, {Soto}, {Jenkins}, {Vu{\v{c}}kovi{\'c}},
  {Ciceri}, \& {Sarkis}}]{jones2018}
{Jones}, M.~I., {Brahm}, R., {Espinoza}, N., {et~al.} 2018, \aap, 613, A76,
  \dodoi{10.1051/0004-6361/201731478}

\bibitem[{{Khandelwal} {et~al.}(2022){Khandelwal}, {Chaturvedi}, {Chakraborty},
  {Sharma}, {Guenther}, {Persson}, {Fridlund}, {Hatzes}, {Prasad}, {Esposito},
  {Chamarthi}, {Nayak}, {Dishendra}, \& {Howell}}]{khandelwal2022}
{Khandelwal}, A., {Chaturvedi}, P., {Chakraborty}, A., {et~al.} 2022, \mnras,
  509, 3339, \dodoi{10.1093/mnras/stab2970}

\bibitem[{Kipping(2013)}]{kipping2013b}
Kipping, D.~M. 2013, Monthly Notices of the Royal Astronomical Society:
  Letters, 434, L51–L55, \dodoi{10.1093/mnrasl/slt075}

\bibitem[{{Kipping}(2013{\natexlab{a}})}]{kipping2013}
{Kipping}, D.~M. 2013{\natexlab{a}}, mnras, 435, 2152,
  \dodoi{10.1093/mnras/stt1435}

\bibitem[{{Kipping}(2013{\natexlab{b}})}]{exoplanet:kipping13}
---. 2013{\natexlab{b}}, \mnras, 435, 2152, \dodoi{10.1093/mnras/stt1435}

\bibitem[{{Kolbl} {et~al.}(2015){Kolbl}, {Marcy}, {Isaacson}, \&
  {Howard}}]{kolbl2015}
{Kolbl}, R., {Marcy}, G.~W., {Isaacson}, H., \& {Howard}, A.~W. 2015, \aj, 149,
  18, \dodoi{10.1088/0004-6256/149/1/18}

\bibitem[{{Kunimoto} {et~al.}(2022){Kunimoto}, {Winn}, {Ricker}, \&
  {Vanderspek}}]{kunimoto2022}
{Kunimoto}, M., {Winn}, J.~N., {Ricker}, G.~R., \& {Vanderspek}, R. 2022, arXiv
  e-prints, arXiv:2202.03656.
\newblock \doarXiv{2202.03656}

\bibitem[{{Li} {et~al.}(2019){Li}, {Tenenbaum}, {Twicken}, {Burke}, {Jenkins},
  {Quintana}, {Rowe}, \& {Seader}}]{li2019}
{Li}, J., {Tenenbaum}, P., {Twicken}, J.~D., {et~al.} 2019, \pasp, 131, 024506,
  \dodoi{10.1088/1538-3873/aaf44d}

\bibitem[{{Lightkurve Collaboration} {et~al.}(2018){Lightkurve Collaboration},
  {Cardoso}, {Hedges}, {Gully-Santiago}, {Saunders}, {Cody}, {Barclay}, {Hall},
  {Sagear}, {Turtelboom}, {Zhang}, {Tzanidakis}, {Mighell}, {Coughlin}, {Bell},
  {Berta-Thompson}, {Williams}, {Dotson}, \& {Barentsen}}]{lightkurve}
{Lightkurve Collaboration}, {Cardoso}, J.~V.~d.~M., {Hedges}, C., {et~al.}
  2018, {Lightkurve: Kepler and TESS time series analysis in Python},
  Astrophysics Source Code Library.
\newblock \doeprint{1812.013}

\bibitem[{{Lillo-Box} {et~al.}(2014){Lillo-Box}, {Barrado}, {Moya},
  {Montesinos}, {Montalb{\'a}n}, {Bayo}, {Barbieri}, {R{\'e}gulo}, {Mancini},
  {Bouy}, \& {Henning}}]{lillobox2014}
{Lillo-Box}, J., {Barrado}, D., {Moya}, A., {et~al.} 2014, \aap, 562, A109,
  \dodoi{10.1051/0004-6361/201322001}

\bibitem[{{Lopez} \& {Fortney}(2016)}]{lopez2016}
{Lopez}, E.~D., \& {Fortney}, J.~J. 2016, \apj, 818, 4,
  \dodoi{10.3847/0004-637X/818/1/4}

\bibitem[{{Luger} {et~al.}(2019){Luger}, {Agol}, {Foreman-Mackey}, {Fleming},
  {Lustig-Yaeger}, \& {Deitrick}}]{exoplanet:luger18}
{Luger}, R., {Agol}, E., {Foreman-Mackey}, D., {et~al.} 2019, aj, 157, 64,
  \dodoi{10.3847/1538-3881/aae8e5}

\bibitem[{{MacLeod} {et~al.}(2018){MacLeod}, {Cantiello}, \&
  {Soares-Furtado}}]{macleod2018}
{MacLeod}, M., {Cantiello}, M., \& {Soares-Furtado}, M. 2018, \apjl, 853, L1,
  \dodoi{10.3847/2041-8213/aaa5fa}

\bibitem[{{Malmquist}(1922)}]{malmquist1922}
{Malmquist}, K.~G. 1922, Meddelanden fran Lunds Astronomiska Observatorium
  Serie I, 100, 1

\bibitem[{{Montalto} {et~al.}(2022){Montalto}, {Malavolta}, {Gregorio},
  {Mantovan}, {Desidera}, {Piotto}, {Nascimbeni}, {Granata}, {Manthopoulou}, \&
  {Claudi}}]{montalto2022}
{Montalto}, M., {Malavolta}, L., {Gregorio}, J., {et~al.} 2022, \mnras, 509,
  2908, \dodoi{10.1093/mnras/stab2923}

\bibitem[{{Naoz}(2016)}]{naoz2016}
{Naoz}, S. 2016, \araa, 54, 441, \dodoi{10.1146/annurev-astro-081915-023315}

\bibitem[{{Nelson} \& {Davis}(1972)}]{nelson1972}
{Nelson}, B., \& {Davis}, W.~D. 1972, \apj, 174, 617, \dodoi{10.1086/151524}

\bibitem[{{Nielsen} {et~al.}(2019){Nielsen}, {Bouchy}, {Turner}, {Giles},
  {Mascare{\~n}o}, {Lovis}, {Marmier}, {Pepe}, {S{\'e}gransan}, {Udry},
  {Otegi}, {Ottoni}, {Stalport}, {Ricker}, {Vanderspek}, {Latham}, {Seager},
  {Winn}, {Jenkins}, {Kane}, {Wittenmyer}, {Bowler}, {Crossfield}, {Horner},
  {Kielkopf}, {Morton}, {Plavchan}, {Tinney}, {Zhang}, {Wright}, {Mengel},
  {Clark}, {Okumura}, {Addison}, {Caldwell}, {Cartwright}, {Collins},
  {Francis}, {Guerrero}, {Huang}, {Matthews}, {Pepper}, {Rose},
  {Villase{\~n}or}, {Wohler}, {Stassun}, {Howell}, {Ciardi}, {Gonzales},
  {Matson}, {Beichman}, \& {Schlieder}}]{nielsen2019}
{Nielsen}, L.~D., {Bouchy}, F., {Turner}, O., {et~al.} 2019, \aap, 623, A100,
  \dodoi{10.1051/0004-6361/201834577}

\bibitem[{{Otor} {et~al.}(2016){Otor}, {Montet}, {Johnson}, {Charbonneau},
  {Collier-Cameron}, {Howard}, {Isaacson}, {Latham}, {Lopez-Morales}, {Lovis},
  {Mayor}, {Micela}, {Molinari}, {Pepe}, {Piotto}, {Phillips}, {Queloz},
  {Rice}, {Sasselov}, {S{\'e}gransan}, {Sozzetti}, {Udry}, \&
  {Watson}}]{otor2016}
{Otor}, O.~J., {Montet}, B.~T., {Johnson}, J.~A., {et~al.} 2016, \aj, 152, 165,
  \dodoi{10.3847/0004-6256/152/6/165}

\bibitem[{{Paxton} {et~al.}(2011){Paxton}, {Bildsten}, {Dotter}, {Herwig},
  {Lesaffre}, \& {Timmes}}]{paxton2011}
{Paxton}, B., {Bildsten}, L., {Dotter}, A., {et~al.} 2011, \apjs, 192, 3,
  \dodoi{10.1088/0067-0049/192/1/3}

\bibitem[{{Peacock}(1983)}]{peacock1983}
{Peacock}, J.~A. 1983, \mnras, 202, 615, \dodoi{10.1093/mnras/202.3.615}

\bibitem[{{Petigura}(2015)}]{petigura2015}
{Petigura}, E.~A. 2015, PhD thesis, University of California, Berkeley

\bibitem[{{Petigura} {et~al.}(2018){Petigura}, {Marcy}, {Winn}, {Weiss},
  {Fulton}, {Howard}, {Sinukoff}, {Isaacson}, {Morton}, \&
  {Johnson}}]{petigura2018}
{Petigura}, E.~A., {Marcy}, G.~W., {Winn}, J.~N., {et~al.} 2018, \aj, 155, 89,
  \dodoi{10.3847/1538-3881/aaa54c}

\bibitem[{{Petrovich}(2015{\natexlab{a}})}]{petrovich2015a}
{Petrovich}, C. 2015{\natexlab{a}}, \apj, 805, 75,
  \dodoi{10.1088/0004-637X/805/1/75}

\bibitem[{{Petrovich}(2015{\natexlab{b}})}]{petrovich2015b}
---. 2015{\natexlab{b}}, \apj, 799, 27, \dodoi{10.1088/0004-637X/799/1/27}

\bibitem[{Press {et~al.}(2007)Press, Teukolsky, Vetterling, \&
  Flannery}]{press2007}
Press, W.~H., Teukolsky, S.~A., Vetterling, W.~T., \& Flannery, B.~P. 2007,
  Numerical recipes 3rd edition: The art of scientific computing (Cambridge
  university press)

\bibitem[{{Price-Whelan} {et~al.}(2018){Price-Whelan}, {Sip{\\H{o}}cz},
  {G{\\"u}nther}, {Lim}, {Crawford}, {Conseil}, {Shupe}, {Craig}, {Dencheva},
  {Ginsburg}, {VanderPlas}, {Bradley}, {P{\\'e}rez-Su{\\'a}rez}, {de
  Val-Borro}, {Paper Contributors}, {Aldcroft}, {Cruz}, {Robitaille},
  {Tollerud}, {Coordination Committee}, {Ardelean}, {Babej}, {Bach},
  {Bachetti}, {Bakanov}, {Bamford}, {Barentsen}, {Barmby}, {Baumbach}, {Berry},
  {Biscani}, {Boquien}, {Bostroem}, {Bouma}, {Brammer}, {Bray}, {Breytenbach},
  {Buddelmeijer}, {Burke}, {Calderone}, {Cano Rodr{\\'\\i}guez}, {Cara},
  {Cardoso}, {Cheedella}, {Copin}, {Corrales}, {Crichton},
  {D{\\textquoteright}Avella}, {Deil}, {Depagne}, {Dietrich}, {Donath},
  {Droettboom}, {Earl}, {Erben}, {Fabbro}, {Ferreira}, {Finethy}, {Fox},
  {Garrison}, {Gibbons}, {Goldstein}, {Gommers}, {Greco}, {Greenfield},
  {Groener}, {Grollier}, {Hagen}, {Hirst}, {Homeier}, {Horton}, {Hosseinzadeh},
  {Hu}, {Hunkeler}, {Ivezi{\\'c}}, {Jain}, {Jenness}, {Kanarek}, {Kendrew},
  {Kern}, {Kerzendorf}, {Khvalko}, {King}, {Kirkby}, {Kulkarni}, {Kumar},
  {Lee}, {Lenz}, {Littlefair}, {Ma}, {Macleod}, {Mastropietro}, {McCully},
  {Montagnac}, {Morris}, {Mueller}, {Mumford}, {Muna}, {Murphy}, {Nelson},
  {Nguyen}, {Ninan}, {N{\\"o}the}, {Ogaz}, {Oh}, {Parejko}, {Parley},
  {Pascual}, {Patil}, {Patil}, {Plunkett}, {Prochaska}, {Rastogi}, {Reddy
  Janga}, {Sabater}, {Sakurikar}, {Seifert}, {Sherbert}, {Sherwood-Taylor},
  {Shih}, {Sick}, {Silbiger}, {Singanamalla}, {Singer}, {Sladen}, {Sooley},
  {Sornarajah}, {Streicher}, {Teuben}, {Thomas}, {Tremblay}, {Turner},
  {Terr{\\'o}n}, {van Kerkwijk}, {de la Vega}, {Watkins}, {Weaver}, {Whitmore},
  {Woillez}, {Zabalza}, \& {Contributors}}]{astropy2018}
{Price-Whelan}, A.~M., {Sip{\\H{o}}cz}, B.~M., {G{\\"u}nther}, H.~M., {et~al.}
  2018, \\aj, 156, 123, \dodoi{10.3847/1538-3881/aabc4f}

\bibitem[{{Quinn} {et~al.}(2015){Quinn}, {White}, {Latham}, {Chaplin},
  {Handberg}, {Huber}, {Kipping}, {Payne}, {Jiang}, {Silva Aguirre}, {Stello},
  {Sliski}, {Ciardi}, {Buchhave}, {Bedding}, {Davies}, {Hekker}, {Kjeldsen},
  {Kuszlewicz}, {Everett}, {Howell}, {Basu}, {Campante},
  {Christensen-Dalsgaard}, {Elsworth}, {Karoff}, {Kawaler}, {Lund},
  {Lundkvist}, {Esquerdo}, {Calkins}, \& {Berlind}}]{quinn2015}
{Quinn}, S.~N., {White}, T.~R., {Latham}, D.~W., {et~al.} 2015, \apj, 803, 49,
  \dodoi{10.1088/0004-637X/803/2/49}

\bibitem[{{Rauer} {et~al.}(2014){Rauer}, {Catala}, {Aerts}, {Appourchaux},
  {Benz}, {Brandeker}, {Christensen-Dalsgaard}, {Deleuil}, {Gizon}, {Goupil},
  {G{\"u}del}, {Janot-Pacheco}, {Mas-Hesse}, {Pagano}, {Piotto}, {Pollacco},
  {Santos}, {Smith}, {Su{\'a}rez}, {Szab{\'o}}, {Udry}, {Adibekyan}, {Alibert},
  {Almenara}, {Amaro-Seoane}, {Eiff}, {Asplund}, {Antonello}, {Barnes},
  {Baudin}, {Belkacem}, {Bergemann}, {Bihain}, {Birch}, {Bonfils}, {Boisse},
  {Bonomo}, {Borsa}, {Brand{\~a}o}, {Brocato}, {Brun}, {Burleigh}, {Burston},
  {Cabrera}, {Cassisi}, {Chaplin}, {Charpinet}, {Chiappini}, {Church},
  {Csizmadia}, {Cunha}, {Damasso}, {Davies}, {Deeg}, {D{\'\i}az}, {Dreizler},
  {Dreyer}, {Eggenberger}, {Ehrenreich}, {Eigm{\"u}ller}, {Erikson}, {Farmer},
  {Feltzing}, {de Oliveira Fialho}, {Figueira}, {Forveille}, {Fridlund},
  {Garc{\'\i}a}, {Giommi}, {Giuffrida}, {Godolt}, {Gomes da Silva}, {Granzer},
  {Grenfell}, {Grotsch-Noels}, {G{\"u}nther}, {Haswell}, {Hatzes},
  {H{\'e}brard}, {Hekker}, {Helled}, {Heng}, {Jenkins}, {Johansen},
  {Khodachenko}, {Kislyakova}, {Kley}, {Kolb}, {Krivova}, {Kupka}, {Lammer},
  {Lanza}, {Lebreton}, {Magrin}, {Marcos-Arenal}, {Marrese}, {Marques},
  {Martins}, {Mathis}, {Mathur}, {Messina}, {Miglio}, {Montalban}, {Montalto},
  {Monteiro}, {Moradi}, {Moravveji}, {Mordasini}, {Morel}, {Mortier},
  {Nascimbeni}, {Nelson}, {Nielsen}, {Noack}, {Norton}, {Ofir}, {Oshagh},
  {Ouazzani}, {P{\'a}pics}, {Parro}, {Petit}, {Plez}, {Poretti}, {Quirrenbach},
  {Ragazzoni}, {Raimondo}, {Rainer}, {Reese}, {Redmer}, {Reffert},
  {Rojas-Ayala}, {Roxburgh}, {Salmon}, {Santerne}, {Schneider}, {Schou},
  {Schuh}, {Schunker}, {Silva-Valio}, {Silvotti}, {Skillen}, {Snellen}, {Sohl},
  {Sousa}, {Sozzetti}, {Stello}, {Strassmeier}, {{\v{S}}vanda}, {Szab{\'o}},
  {Tkachenko}, {Valencia}, {Van Grootel}, {Vauclair}, {Ventura}, {Wagner},
  {Walton}, {Weingrill}, {Werner}, {Wheatley}, \& {Zwintz}}]{rauer2014}
{Rauer}, H., {Catala}, C., {Aerts}, C., {et~al.} 2014, Experimental Astronomy,
  38, 249, \dodoi{10.1007/s10686-014-9383-4}

\bibitem[{{Reffert} {et~al.}(2015){Reffert}, {Bergmann}, {Quirrenbach},
  {Trifonov}, \& {K{\"u}nstler}}]{reffert2015}
{Reffert}, S., {Bergmann}, C., {Quirrenbach}, A., {Trifonov}, T., \&
  {K{\"u}nstler}, A. 2015, \aap, 574, A116, \dodoi{10.1051/0004-6361/201322360}

\bibitem[{Ricker {et~al.}(2014)Ricker, Winn, Vanderspek, Latham, Bakos, Bean,
  Berta-Thompson, Brown, Buchhave, Butler, Butler, Chaplin, Charbonneau,
  Christensen-Dalsgaard, Clampin, Deming, Doty, Lee, Dressing, Dunham, Endl,
  Fressin, Ge, Henning, Holman, Howard, Ida, Jenkins, Jernigan, Johnson,
  Kaltenegger, Kawai, Kjeldsen, Laughlin, Levine, Lin, Lissauer, MacQueen,
  Marcy, McCullough, Morton, Narita, Paegert, Palle, Pepe, Pepper, Quirrenbach,
  Rinehart, Sasselov, Sato, Seager, Sozzetti, Stassun, Sullivan, Szentgyorgyi,
  Torres, Udry, \& Villasenor}]{ricker2014}
Ricker, G.~R., Winn, J.~N., Vanderspek, R., {et~al.} 2014, Journal of
  Astronomical Telescopes, Instruments, and Systems, 1, 1 ,
  \dodoi{10.1117/1.JATIS.1.1.014003}

\bibitem[{{Rodriguez} {et~al.}(2021){Rodriguez}, {Quinn}, {Zhou}, {Vanderburg},
  {Nielsen}, {Wittenmyer}, {Brahm}, {Reed}, {Huang}, {Vach}, {Ciardi},
  {Oelkers}, {Stassun}, {Hellier}, {Gaudi}, {Eastman}, {Collins}, {Bieryla},
  {Christian}, {Latham}, {Carleo}, {Wright}, {Matthews}, {Gonzales}, {Ziegler},
  {Dressing}, {Howell}, {Tan}, {Wittrock}, {Plavchan}, {McLeod}, {Baker},
  {Wang}, {Radford}, {Schwarz}, {Esposito}, {Ricker}, {Vanderspek}, {Seager},
  {Winn}, {Jenkins}, {Addison}, {Anderson}, {Barclay}, {Beatty}, {Berlind},
  {Bouchy}, {Bowen}, {Bowler}, {Brasseur}, {Brice{\~n}o}, {Caldwell},
  {Calkins}, {Cartwright}, {Chaturvedi}, {Chaverot}, {Chimaladinne},
  {Christiansen}, {Collins}, {Crossfield}, {Eastridge}, {Espinoza}, {Esquerdo},
  {Feliz}, {Fenske}, {Fong}, {Gan}, {Giacalone}, {Gill}, {Gordon}, {Granados},
  {Grieves}, {Guenther}, {Guerrero}, {Henning}, {Henze}, {Hesse}, {Hobson},
  {Horner}, {James}, {Jensen}, {Jimenez}, {Jord{\'a}n}, {Kane}, {Kielkopf},
  {Kim}, {Kuhn}, {Latouf}, {Law}, {Levine}, {Lund}, {Mann}, {Mao}, {Matson},
  {Mengel}, {Mink}, {Newman}, {O'Dwyer}, {Okumura}, {Palle}, {Pepper},
  {Quintana}, {Sarkis}, {Savel}, {Schlieder}, {Schnaible}, {Shporer}, {Sefako},
  {Seidel}, {Siverd}, {Skinner}, {Stalport}, {Stevens}, {Stibbards}, {Tinney},
  {West}, {Yahalomi}, \& {Zhang}}]{rodriguez2021}
{Rodriguez}, J.~E., {Quinn}, S.~N., {Zhou}, G., {et~al.} 2021, \aj, 161, 194,
  \dodoi{10.3847/1538-3881/abe38a}

\bibitem[{{Safonov} {et~al.}(2017){Safonov}, {Lysenko}, \&
  {Dodin}}]{safonov2017}
{Safonov}, B.~S., {Lysenko}, P.~A., \& {Dodin}, A.~V. 2017, Astronomy Letters,
  43, 344, \dodoi{10.1134/S1063773717050036}

\bibitem[{Salvatier {et~al.}(2016)Salvatier, Wiecki, \&
  Fonnesbeck}]{exoplanet:pymc3}
Salvatier, J., Wiecki, T.~V., \& Fonnesbeck, C. 2016, PeerJ Computer Science,
  2, e55

\bibitem[{{Santos} {et~al.}(2021){Santos}, {Breton}, {Mathur}, \&
  {Garc{\'\i}a}}]{santos2021}
{Santos}, A.~R.~G., {Breton}, S.~N., {Mathur}, S., \& {Garc{\'\i}a}, R.~A.
  2021, \apjs, 255, 17, \dodoi{10.3847/1538-4365/ac033f}

\bibitem[{{Saunders} {et~al.}(2022){Saunders}, {Grunblatt}, {Huber}, {Collins},
  {Jensen}, {Vanderburg}, {Brahm}, {Jord{\'a}n}, {Espinoza}, {Henning},
  {Hobson}, {Quinn}, {Zhou}, {Butler}, {Crause}, {Kuhn}, {Moses Mogotsi},
  {Hellier}, {Angus}, {Hattori}, {Chontos}, {Ricker}, {Jenkins}, {Tenenbaum},
  {Latham}, {Seager}, {Vanderspek}, {Winn}, {Stockdale}, \&
  {Cloutier}}]{saunders2022}
{Saunders}, N., {Grunblatt}, S.~K., {Huber}, D., {et~al.} 2022, \aj, 163, 53,
  \dodoi{10.3847/1538-3881/ac38a1}

\bibitem[{{Schlaufman} \& {Winn}(2013)}]{schlaufman2013}
{Schlaufman}, K.~C., \& {Winn}, J.~N. 2013, \apj, 772, 143,
  \dodoi{10.1088/0004-637X/772/2/143}

\bibitem[{{Scott} {et~al.}(2018){Scott}, {Howell}, {Horch}, \&
  {Everett}}]{scott2018}
{Scott}, N.~J., {Howell}, S.~B., {Horch}, E.~P., \& {Everett}, M.~E. 2018,
  \pasp, 130, 054502, \dodoi{10.1088/1538-3873/aab484}

\bibitem[{{Smith} {et~al.}(2017){Smith}, {Gandolfi}, {Barrag{\'a}n}, {Bowler},
  {Csizmadia}, {Endl}, {Fridlund}, {Grziwa}, {Guenther}, {Hatzes}, {Nowak},
  {Albrecht}, {Alonso}, {Cabrera}, {Cochran}, {Deeg}, {Cusano},
  {Eigm{\"u}ller}, {Erikson}, {Hidalgo}, {Hirano}, {Johnson}, {Korth}, {Mann},
  {Narita}, {Nespral}, {Palle}, {P{\"a}tzold}, {Prieto-Arranz}, {Rauer},
  {Ribas}, {Tingley}, \& {Wolthoff}}]{smith2017}
{Smith}, A.~M.~S., {Gandolfi}, D., {Barrag{\'a}n}, O., {et~al.} 2017, \mnras,
  464, 2708, \dodoi{10.1093/mnras/stw2487}

\bibitem[{{Stassun} {et~al.}(2019){Stassun}, {Oelkers}, {Paegert}, {Torres},
  {Pepper}, {De Lee}, {Collins}, {Latham}, {Muirhead}, {Chittidi},
  {Rojas-Ayala}, {Fleming}, {Rose}, {Tenenbaum}, {Ting}, {Kane}, {Barclay},
  {Bean}, {Brassuer}, {Charbonneau}, {Ge}, {Lissauer}, {Mann}, {McLean},
  {Mullally}, {Narita}, {Plavchan}, {Ricker}, {Sasselov}, {Seager}, {Sharma},
  {Shiao}, {Sozzetti}, {Stello}, {Vanderspek}, {Wallace}, \&
  {Winn}}]{stassun2019}
{Stassun}, K.~G., {Oelkers}, R.~J., {Paegert}, M., {et~al.} 2019, \aj, 158,
  138, \dodoi{10.3847/1538-3881/ab3467}

\bibitem[{{Sullivan} {et~al.}(2015){Sullivan}, {Winn}, {Berta-Thompson},
  {Charbonneau}, {Deming}, {Dressing}, {Latham}, {Levine}, {McCullough},
  {Morton}, {Ricker}, {Vanderspek}, \& {Woods}}]{sullivan2015}
{Sullivan}, P.~W., {Winn}, J.~N., {Berta-Thompson}, Z.~K., {et~al.} 2015, \apj,
  809, 77, \dodoi{10.1088/0004-637X/809/1/77}

\bibitem[{{Sun} {et~al.}(2018){Sun}, {Arras}, {Weinberg}, {Troup}, \&
  {Majewski}}]{sun2018}
{Sun}, M., {Arras}, P., {Weinberg}, N.~N., {Troup}, N.~W., \& {Majewski}, S.~R.
  2018, \mnras, 481, 4077, \dodoi{10.1093/mnras/sty2464}

\bibitem[{{Takarada} {et~al.}(2018){Takarada}, {Sato}, {Omiya}, {Harakawa},
  {Nagasawa}, {Izumiura}, {Kambe}, {Takeda}, {Yoshida}, {Itoh}, {Ando},
  {Kokubo}, \& {Ida}}]{takarada2018}
{Takarada}, T., {Sato}, B., {Omiya}, M., {et~al.} 2018, \pasj, 70, 59,
  \dodoi{10.1093/pasj/psy052}

\bibitem[{Tange(2018)}]{tange2018}
Tange, O. 2018, GNU Parallel 2018 (Ole Tange), \dodoi{10.5281/zenodo.1146014}

\bibitem[{{Tayar} {et~al.}(2022){Tayar}, {Claytor}, {Huber}, \& {van
  Saders}}]{tayar2022}
{Tayar}, J., {Claytor}, Z.~R., {Huber}, D., \& {van Saders}, J. 2022, \apj,
  927, 31, \dodoi{10.3847/1538-4357/ac4bbc}

\bibitem[{{Theano Development Team}(2016)}]{exoplanet:theano}
{Theano Development Team}. 2016, arXiv e-prints, abs/1605.02688.
\newblock \url{http://arxiv.org/abs/1605.02688}

\bibitem[{{Turner} {et~al.}(2021){Turner}, {Ridden-Harper}, \&
  {Jayawardhana}}]{turner2021}
{Turner}, J.~D., {Ridden-Harper}, A., \& {Jayawardhana}, R. 2021, \aj, 161, 72,
  \dodoi{10.3847/1538-3881/abd178}

\bibitem[{{Twicken} {et~al.}(2018){Twicken}, {Catanzarite}, {Clarke},
  {Girouard}, {Jenkins}, {Klaus}, {Li}, {McCauliff}, {Seader}, {Tenenbaum},
  {Wohler}, {Bryson}, {Burke}, {Caldwell}, {Haas}, {Henze}, \&
  {Sanderfer}}]{twicken2018}
{Twicken}, J.~D., {Catanzarite}, J.~H., {Clarke}, B.~D., {et~al.} 2018, \pasp,
  130, 064502, \dodoi{10.1088/1538-3873/aab694}

\bibitem[{{Ulmer-Moll} {et~al.}(2022){Ulmer-Moll}, {Lendl}, {Gill},
  {Villanueva}, {Hobson}, {Bouchy}, {Brahm}, {Dragomir}, {Grieves},
  {Mordasini}, {Anderson}, {Acton}, {Bayliss}, {Bieryla}, {Burleigh},
  {Casewell}, {Chaverot}, {Eigm{\"u}ller}, {Feliz}, {Gaudi}, {Gillen}, {Goad},
  {Gupta}, {G{\"u}nther}, {Henderson}, {Henning}, {Jenkins}, {Jones},
  {Jord{\'a}n}, {Kendall}, {Latham}, {Mireles}, {Moyano}, {Nadol}, {Osborn},
  {Pepper}, {Pinto}, {Psaridi}, {Queloz}, {Quinn}, {Rojas}, {Sarkis},
  {Schlecker}, {Tilbrook}, {Torres}, {Trifonov}, {Udry}, {Vines}, {West},
  {Wheatley}, {Yao}, {Zhao}, \& {Zhou}}]{ulmermoll2022}
{Ulmer-Moll}, S., {Lendl}, M., {Gill}, S., {et~al.} 2022, \aap, 666, A46,
  \dodoi{10.1051/0004-6361/202243583}

\bibitem[{{Van Eylen} {et~al.}(2016){Van Eylen}, {Albrecht}, {Gandolfi}, {Dai},
  {Winn}, {Hirano}, {Narita}, {Bruntt}, {Prieto-Arranz}, {Bejar}, {Nowak},
  {Lund}, {Palle}, {Ribas}, {Sanchis-Ojeda}, {Yu}, {Arriagada}, {Butler},
  {Crane}, {Handberg}, {Deeg}, {Jessen-Hansen}, {Johnson}, {Nespral}, {Rogers},
  {Ryu}, {Shectman}, {Shrotriya}, {Slumstrup}, {Takeda}, {Teske}, {Thompson},
  {Vanderburg}, \& {Wittenmyer}}]{vaneylen2016}
{Van Eylen}, V., {Albrecht}, S., {Gandolfi}, D., {et~al.} 2016, ArXiv e-prints.
\newblock \doarXiv{1605.09180}

\bibitem[{{Van Eylen} {et~al.}(2018){Van Eylen}, {Dai}, {Mathur}, {Gandolfi},
  {Albrecht}, {Fridlund}, {Garc{\'\i}a}, {Guenther}, {Hjorth}, {Justesen},
  {Livingston}, {Lund}, {P{\'e}rez Hern{\'a}ndez}, {Prieto-Arranz}, {Regulo},
  {Bugnet}, {Everett}, {Hirano}, {Nespral}, {Nowak}, {Palle}, {Silva Aguirre},
  {Trifonov}, {Winn}, {Barrag{\'a}n}, {Beck}, {Chaplin}, {Cochran},
  {Csizmadia}, {Deeg}, {Endl}, {Heeren}, {Grziwa}, {Hatzes}, {Hidalgo},
  {Korth}, {Mathis}, {Monta{\~n}es Rodriguez}, {Narita}, {Patzold}, {Persson},
  {Rodler}, \& {Smith}}]{vaneylen2018}
{Van Eylen}, V., {Dai}, F., {Mathur}, S., {et~al.} 2018, \mnras, 478, 4866,
  \dodoi{10.1093/mnras/sty1390}

\bibitem[{Van~Eylen {et~al.}(2019)Van~Eylen, Albrecht, Huang, MacDonald,
  Dawson, Cai, Foreman-Mackey, Lundkvist, Aguirre, Snellen, \&
  et~al.}]{vaneylen2019}
Van~Eylen, V., Albrecht, S., Huang, X., {et~al.} 2019, The Astronomical
  Journal, 157, 61, \dodoi{10.3847/1538-3881/aaf22f}

\bibitem[{{Veras} {et~al.}(2015){Veras}, {Brown}, {Mustill}, \&
  {Pollacco}}]{veras2015}
{Veras}, D., {Brown}, D. J.~A., {Mustill}, A.~J., \& {Pollacco}, D. 2015,
  \mnras, 453, 67, \dodoi{10.1093/mnras/stv1615}

\bibitem[{{Veras} {et~al.}(2013){Veras}, {Mustill}, {Bonsor}, \&
  {Wyatt}}]{veras2013}
{Veras}, D., {Mustill}, A.~J., {Bonsor}, A., \& {Wyatt}, M.~C. 2013, \mnras,
  431, 1686, \dodoi{10.1093/mnras/stt289}

\bibitem[{{Villaver} \& {Livio}(2009)}]{villaver2009}
{Villaver}, E., \& {Livio}, M. 2009, \apjl, 705, L81,
  \dodoi{10.1088/0004-637X/705/1/L81}

\bibitem[{{Villaver} {et~al.}(2014){Villaver}, {Livio}, {Mustill}, \&
  {Siess}}]{villaver2014}
{Villaver}, E., {Livio}, M., {Mustill}, A.~J., \& {Siess}, L. 2014, \apj, 794,
  3, \dodoi{10.1088/0004-637X/794/1/3}

\bibitem[{{Vogt} {et~al.}(1994){Vogt}, {Allen}, {Bigelow}, {Bresee}, {Brown},
  {Cantrall}, {Conrad}, {Couture}, {Delaney}, {Epps}, {Hilyard}, {Hilyard},
  {Horn}, {Jern}, {Kanto}, {Keane}, {Kibrick}, {Lewis}, {Osborne},
  {Pardeilhan}, {Pfister}, {Ricketts}, {Robinson}, {Stover}, {Tucker}, {Ward},
  \& {Wei}}]{vogt1994}
{Vogt}, S.~S., {Allen}, S.~L., {Bigelow}, B.~C., {et~al.} 1994, in \procspie,
  Vol. 2198, Instrumentation in Astronomy VIII, ed. D.~L. {Crawford} \& E.~R.
  {Craine}, 362

\bibitem[{{Wang} {et~al.}(2019){Wang}, {Jones}, {Shporer}, {Fulton}, {Paredes},
  {Trifonov}, {Kossakowski}, {Eastman}, {Redfield}, {G{\"u}nther}, {Kreidberg},
  {Huang}, {Millholland}, {Seligman}, {Fischer}, {Brahm}, {Wang}, {Cruz},
  {Henry}, {James}, {Addison}, {Liang}, {Davis}, {Tronsgaard}, {Worku},
  {Brewer}, {K{\"u}rster}, {Zhang}, {Beichman}, {Bieryla}, {Brown},
  {Christiansen}, {Ciardi}, {Collins}, {Esquerdo}, {Howard}, {Isaacson},
  {Latham}, {Mazeh}, {Petigura}, {Quinn}, {Shahaf}, {Siverd}, {Rodler},
  {Reffert}, {Zakhozhay}, {Ricker}, {Vanderspek}, {Seager}, {Winn}, {Jenkins},
  {Boyd}, {F{\H{u}}r{\'e}sz}, {Henze}, {Levine}, {Morris}, {Paegert},
  {Stassun}, {Ting}, {Vezie}, \& {Laughlin}}]{wang2019}
{Wang}, S., {Jones}, M., {Shporer}, A., {et~al.} 2019, \aj, 157, 51,
  \dodoi{10.3847/1538-3881/aaf1b7}

\bibitem[{{Wittenmyer} {et~al.}(2022){Wittenmyer}, {Clark}, {Trifonov},
  {Addison}, {Wright}, {Stassun}, {Horner}, {Lowson}, {Kielkopf}, {Kane},
  {Plavchan}, {Shporer}, {Zhang}, {Bowler}, {Mengel}, {Okumura}, {Rabus},
  {Johnson}, {Harbeck}, {Tronsgaard}, {Buchhave}, {Collins}, {Collins}, {Gan},
  {Jensen}, {Howell}, {Furlan}, {Gnilka}, {Lester}, {Matson}, {Scott},
  {Ricker}, {Vanderspek}, {Latham}, {Seager}, {Winn}, {Jenkins}, {Rudat},
  {Quintana}, {Rodriguez}, {Caldwell}, {Quinn}, {Essack}, \&
  {Bouma}}]{wittenmyer2022}
{Wittenmyer}, R.~A., {Clark}, J.~T., {Trifonov}, T., {et~al.} 2022, \aj, 163,
  82, \dodoi{10.3847/1538-3881/ac3f39}

\bibitem[{{Wolthoff} {et~al.}(2022){Wolthoff}, {Reffert}, {Quirrenbach},
  {Jones}, {Wittenmyer}, \& {Jenkins}}]{wolthoff2022}
{Wolthoff}, V., {Reffert}, S., {Quirrenbach}, A., {et~al.} 2022, arXiv
  e-prints, arXiv:2202.12800.
\newblock \doarXiv{2202.12800}

\bibitem[{{Wong} {et~al.}(2022){Wong}, {Shporer}, {Vissapragada},
  {Greklek-McKeon}, {Knutson}, {Winn}, \& {Benneke}}]{wong2022}
{Wong}, I., {Shporer}, A., {Vissapragada}, S., {et~al.} 2022, \aj, 163, 175,
  \dodoi{10.3847/1538-3881/ac5680}

\bibitem[{{Xie} {et~al.}(2016){Xie}, {Dong}, {Zhu}, {Huber}, {Zheng}, {De Cat},
  {Fu}, {Liu}, {Luo}, {Wu}, {Zhang}, {Zhang}, {Zhou}, {Cao}, {Hou}, {Wang}, \&
  {Zhang}}]{xie2016}
{Xie}, J.-W., {Dong}, S., {Zhu}, Z., {et~al.} 2016, Proceedings of the National
  Academy of Science, 113, 11431, \dodoi{10.1073/pnas.1604692113}

\bibitem[{{Yee} {et~al.}(2020){Yee}, {Winn}, {Knutson}, {Patra},
  {Vissapragada}, {Zhang}, {Holman}, {Shporer}, \& {Wright}}]{yee2020}
{Yee}, S.~W., {Winn}, J.~N., {Knutson}, H.~A., {et~al.} 2020, \apjl, 888, L5,
  \dodoi{10.3847/2041-8213/ab5c16}

\bibitem[{{Zahn}(1977)}]{zahn1977}
{Zahn}, J.~P. 1977, \aap, 57, 383

\bibitem[{{Zahn}(1989)}]{1989A&A...220..112Z}
---. 1989, \aap, 220, 112

\end{thebibliography}
\bibliographystyle{aasjournal}

\end{document}